\documentclass[aps,preprint,groupedaddress,superscriptaddress,longbibliography]{revtex4}
\usepackage{lineno}
\usepackage{amssymb,amsfonts,amsmath}
\usepackage{adjustbox}
\usepackage[dvipsnames]{xcolor}
\usepackage{braket}

\usepackage{float}
\usepackage{setspace}
\usepackage{ulem}
\usepackage{graphicx}
\usepackage{array}
\usepackage{subcaption}
\usepackage{soul}

\usepackage{placeins}
\usepackage{anyfontsize}
\usepackage{rotating}

\usepackage{graphicx}% Include figure files
\usepackage{dcolumn}% Align table columns on a decimal point
\usepackage{bm}% bold math
\usepackage{amsmath}%math
\usepackage{commath}%math formatting
\usepackage[dvipsnames]{xcolor}
\usepackage{placeins} % FloatBarrier
\usepackage{dblfloatfix}
\usepackage[utf8]{inputenc}
\usepackage[T1]{fontenc}
\usepackage{mathptmx}
\usepackage{etoolbox}
\usepackage{bbold}
\usepackage{hyperref}
\usepackage[english]{babel}

\usepackage{xr-hyper}
\usepackage{xr}
\begin{document}

\title{Interpretable Graph Neural Networks for Classifying Structure and Magnetism in Delafossite Compounds}

\author{Jovin Ryan Joseph}
\affiliation{Materials Science and Technology Division, Oak Ridge National Laboratory, Oak Ridge, Tennessee 37831, USA}
\affiliation{Artie McFerrin Department of Chemical Engineering, Texas A\&M University,
College Station, Texas 77843, USA}

\author{Do Hoon Kiem}
\affiliation{Materials Science and Technology Division, Oak Ridge National Laboratory, Oak Ridge, Tennessee 37831, USA}
\affiliation{Center for Nanophase Materials Science, Oak Ridge National Laboratory, Oak Ridge, Tennessee 37831, USA}

\author{Sinchul Yeom}
\affiliation{Materials Science and Technology Division, Oak Ridge National Laboratory, Oak Ridge, Tennessee 37831, USA}

\author{Mina Yoon}
\email{myoon@ornl.gov}
\affiliation{Materials Science and Technology Division, Oak Ridge National Laboratory, Oak Ridge, Tennessee 37831, USA}

\begin{abstract}
Delafossites ($ABC_2$, where A and B are metals and C is a chalcogen) are a versatile family of quantum materials and layered oxides/chalcogenides whose properties are highly sensitive to atomic composition and stacking geometry. Their broad chemical tunability makes them an ideal platform for large-scale combinatorial exploration and high-throughput computational screening with desirable quantum properties. In this work, we employ a Concept Whitening Graph Neural Network, a gray-box AI model, to classify delafossite structures by stacking sequence and magnetic states. By aligning learned representations with human-interpretable physical concepts, this graybox approach enables both accurate prediction and insight into the structural/chemical features driving magnetic behavior. The magnetic-ordering models achieved validation accuracies exceeding 80\%, with a further slight uptick observed in the model incorporating the largest number of concepts. Concept alignment analysis revealed measurable learning across nine physically meaningful descriptors, with coefficient of determination values ranging from approximately 0.6 for the $d$-shell valence electron concepts to 0.4–0.5 for the magnetic coupling parameters. Furthermore, we mapped the concept importances onto the material graph representation, elucidating interpretable physical trends and the progression of stable concept-aligned regions across training. These results demonstrate the potential of interpretable graph-based learning to capture the underlying physics of complex materials systems and provide an interpretable framework for accelerating the discovery and understanding of delafossites and related crystalline materials.
\end{abstract}

\onecolumngrid
Notice: This manuscript has been coauthored by UT-Battelle, LLC, under Contract No. DE-AC0500OR22725 with the U.S. Department of Energy. The United States Government retains and the publisher, by accepting the article for publication, acknowledges that the United States Government retains a non-exclusive, paid-up, irrevocable, world-wide license to publish or reproduce the published form of this manuscript, or allow others to do so, for the United States Government purposes. The Department of Energy will provide public access to these results of federally sponsored research in accordance with the DOE Public Access Plan (http://energy.gov/downloads/doe-public-access-plan). 
\newpage

\maketitle

\section{Introduction}
\label{sec:introduction}
Recently, there has been growing interest in the computational materials science community in applying machine learning (ML) and neural networks (NNs) to predict complex material properties and to accelerate the screening of materials for a wide range of technological applications \cite{liu_machine_2023, yang_machine_2023, liao_predicting_2022, fan_accelerate_2024, butler_machine_2018}. Among these approaches, Graph Neural Networks (GNNs) have emerged as a powerful tool for modeling and predicting physical properties \cite{sarada_devi_graph_2024, reiser_graph_2022, gao_molecular_2024, nevolianis_multi-fidelity_2025, rasool_optimizing_2024}. By representing molecules and crystalline solids as graphs, with atoms as nodes and bonds or interactions as edges, GNNs leverage message-passing schemes to achieve greater expressivity, capturing complex structural and relational information that is difficult to model with traditional approaches. However, this increased expressivity comes at the cost of greater model complexity and reduced interpretability. Given the importance of explainable models for reliable scientific discovery, there has been growing interest in developing interpretable neural network architectures or methods to extract meaningful insights from their learned representations. This includes studies specifically in materials science \cite{yuan_explainable_2022, wu_chemistry-intuitive_2023, xie_crystal_2018, banik_evaluating_2024, xiao_graph_2024}, as well as more general approaches across a variety of neural network models \cite{zhong_explainable_2022, yuan_explainability_2023} across diverse fields. While our focus in this work is on GNNs, these broader efforts highlight the importance of interpretability in scientific machine learning.  

One promising approach for improving interpretability in neural networks is Concept Whitening (CW) \cite{chen_concept_2020, proietti_explainable_2024}. CW transforms the latent features of a network into an orthogonalized basis aligned with human-interpretable concepts, allowing researchers to gain insight into how high-level, semantically meaningful information is learned \cite{chen_concept_2020}. Originally developed for image recognition, CW has shown the potential to link predictive performance with interpretability. To our knowledge, its application beyond computer vision has been limited, with only a few examples exploring other domains, such as drug discovery \cite{proietti_explainable_2024}. This suggests an opportunity to explore CW in materials science, where understanding the concepts learned by a model could provide valuable insight into complex material properties.

Delafossite oxides and chalcogenides are an ideal test system for exploring interpretable, graph-based learning. The delafossite framework ($ABC_2$) is structurally diverse and geometrically rich. It consists of octahedral $BC_6$ units and linear $A$-$C$-$A$ chains \cite{marquardt_crystal_2006}. Across this framework. multiple stacking sequesnce - direct, tilt, and crednerite - produce distinct connectivity motifs and graph topologies. 
%Even small variations in layer registry produce significant structural signatures and corresponding graph representations learned by GNNs. 
These structural differences play a central role in determining the physical behavior of delafossites, which exhibit intriguing electronic \cite{noh_anisotropic_2009, benreguia_optical_2024, lunca-popa_tuneable_2020, matasov_3r-cucro2_2024}, optical \cite{taddee_characterization_2016, gharbi_exploring_2025, el-bassuony_influence_2022}, and magnetic \cite{bouda_unexpected_2019, fujii_tlybse2_2025, ketfi_insight_2023} properties. These properties are highly sensitive to the atomic arrangement and stacking sequence. Therefore, an accurate representation of crystal geometry is essential for predictive modeling. Moreover, the delafossite structure is chemically highly accommodating, allowing for numerous combinations of metallic elements and chalcogens. This structural and chemical flexibility opens up a vast combinatorial design space that is well suited for high-throughput screening and AI-driven structure-property discovery. From a physical perspective, delafossites are well-suited for concept-based interpretability because their magnetic behavior stems from local coordination distortions, $d$-electron occupancy, and competing in-plane and out-of-plane magnetic interactions. These quantities can be directly encoded as explicit concepts in a Concept Whitening layer. Consequently, delafossites provide a scientifically meaningful platform in which geometric, electronic, and magnetic descriptors are clearly defined, physically interpretable, and directly linked to target properties. These features make delafossites an excellent benchmark system for evaluating whether gray-box neural network approaches, such as Concept Whitening Graph Neural Networks (CW-GNNs), genuinely learn and align with the underlying physics.
%In this study, we investigate the use of Concept Whitening specifically within GNNs for predicting both the stacking type and magnetic ordering of Delafossite crystal structures. Delafossites, with the general formula $ABC_2$, are a class of layered oxide materials consisting of octahedral $BC_6$ units and linear $A$-$C$-$A$ chains \cite{marquardt_crystal_2006}. These materials exhibit a range of intriguing electronic \cite{noh_anisotropic_2009, benreguia_optical_2024, lunca-popa_tuneable_2020, matasov_3r-cucro2_2024}, optical \cite{taddee_characterization_2016, gharbi_exploring_2025, el-bassuony_influence_2022}, and magnetic \cite{bouda_unexpected_2019, fujii_tlybse2_2025, ketfi_insight_2023} properties. These properties are highly sensitive to the atomic arrangement and stacking sequence within the crystal lattice, making structural information critical for accurate predictions. Moreover, the Delafossite structure is highly accommodating, allowing many variations in atomic composition and stacking, which leads to a combinatorial explosion in the number of possible structures. To evaluate the effectiveness of Concept Whitening in GNNs, we first benchmark its performance against standard batch normalization using a dataset of Delafossite structural variations. We then apply the CW-GNN model to predict magnetic ordering, demonstrating a task that has previously been challenging to achieve with conventional approaches.

This study investigates using Concept Whitening in graph neural networks to interpretably classify both stacking types and magnetic ordering in delafossite family of layered oxides and chalcogenides. First, we benchmark CW-GNNs against standard batch-normalized GNNs using a large dataset of 33,520 relaxed structures. We evaluate the robustness of CW-GNNs with respect to concept-loss weighting, concept noise, label corruption, and dataset size. Next, we extend the approach to a more complex magnetic-ordering task. We incorporate nine physically meaningful descriptors, including $d$-shell electron counts and magnetic exchange-coupling parameters, to assess how well CW aligns latent representations with known physical concepts. The CW-GNN achieves competitive accuracy while providing significantly enhanced interpretability. Furthermore, mapping learned concept importances onto the material graph reveals clear, physics-guided trends and demonstrates the emergence of stable, concept-aligned regions throughout training. Together, these results demonstrate that interpretable, graph-based learning can accurately capture the essential structural and magnetic features of delafossite materials. This approach provides a transparent framework for accelerating the discovery and understanding of related crystalline systems.

\section{Results and discussion}
\label{sec:results_and_discussion}

%\mycom{To Jovin: When you make the edits, please use \textcolor{green}{greenlines} to mark all changes, so I can quickly see exactly what was added or modified - this will save a lot of time.. }
%\FloatBarrier
\subsection{Delafossites - Datasets}
\label{sec:datasets}
\begin{figure}[htbp]
    \centering
    \includegraphics[width=\linewidth]{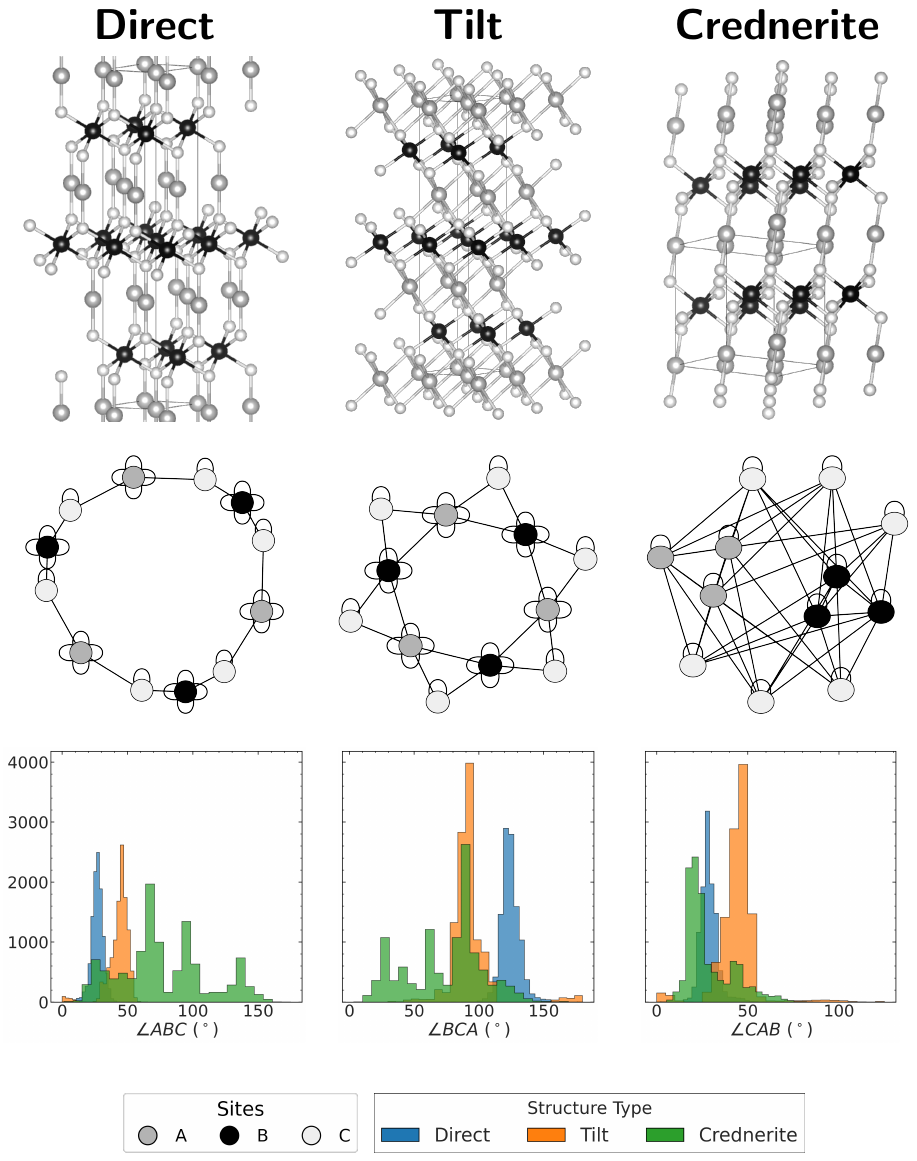}
    \caption{Crystal structures of delafossites with direct, tilt, and crednerite stacking (top row) and their corresponding graph representations (bottom row), generated using the workflow described in Section~\ref{sec:methods_graph_construction}. In the graph representations, atoms are shown as nodes (circles color-coded by site type: A, B, or C) and bonds as edges (lines between nodes). Each node contains one self-loop to encode its own features, with additional self-loops representing connections to its periodic images. %\textcolor{green}{
The bottom row shows histograms of the three atomic triplet angles: (a) $\angle ABC$, (b) $\angle BCA$, and (c) $\angle CAB$, color-coded by stacking type.
%}.
}
    \label{fig:stacking_graph_grid}
\end{figure}

%\subsection{Delafossite Structure Dataset (Benchmarking)}
%\label{sec:datasets_structures}
The primary dataset consists of 33,520 relazed delafossite structures, each following the general formula $ABC_2$, where $A$ and $B$ are metals and $C$ is a chalcogen. Starting from the three canonical delafossite stacking types - direct, tilt, and Crednerite (see Fig.~\ref{fig:stacking_graph_grid}) - we systematically substituted the A and B sites with 57 metallic elements spanning a broad chemical range of the periodic table, including alkali, alkaline-earth, transition, post-transition, and lanthanide elements. This broad selection was chosen to capture the wide range of ionic sizes, valence states, and electronic configurations known to influence delafossite stability. The chalcogen site was substituted with O, S, Se, or Te; structures with identical A and B species were excluded; and both $ABO_2$ and $BAO_2$ permutations were included. All initial structures (38,304 total) were fully optimized (cell shape, cell volume, and atomic positions) using Density Functional Theory (DFT) calculations with the Vienna Ab-initio Simulation Package (VASP)~\cite{kresse_ab_1993, kresse_ab_1994} until force convergence was achieved (see the Methods section for details). We only included converged structures in subsequent analyses, yielding a final dataset of 33,520 relaxed geometries for the stacking-type classification benchmarks.

Figure~\ref{fig:stacking_graph_grid} illustrates the atomic arrangements and corresponding graph representations of the direct, tilt, and Crednerite stacking types. The graphs clearly show the morphological differences among the stacking configurations: Direct stacking produces a compact, circular graph, tilt produces a six-pointed star-like pattern, and crednerite forms a dense, highly interconnected graph with clear separation among the $A$, $B$, and $C$ node clusters. 
%The figure also includes histograms of the geometric triplet-angle distributions, which further emphasize the structural distinctions between stacking types.
These structural differences are also reflected in the $A$-$B$-$C$ triplet angle distributions, shown in the histograms in Fig.~\ref{fig:stacking_graph_grid}. Crednerite stacking exhibits wide variations in angles ($\angle ABC$, $\angle BCA$, and $\angle CAB$) due to significantly distorted coordination environments that disrupt the uniformity of the interlayer geometry. On the other hand, direct and tilt show more constrained and characteristic geometric signatures. Since the stacking arrangement strongly influences physical behavior, including magnetic ordering, these angle-based features serve as interpretable geometric concepts for the CW layer.

\begin{figure}[htbp]
    \centering
    \includegraphics[width=\linewidth]{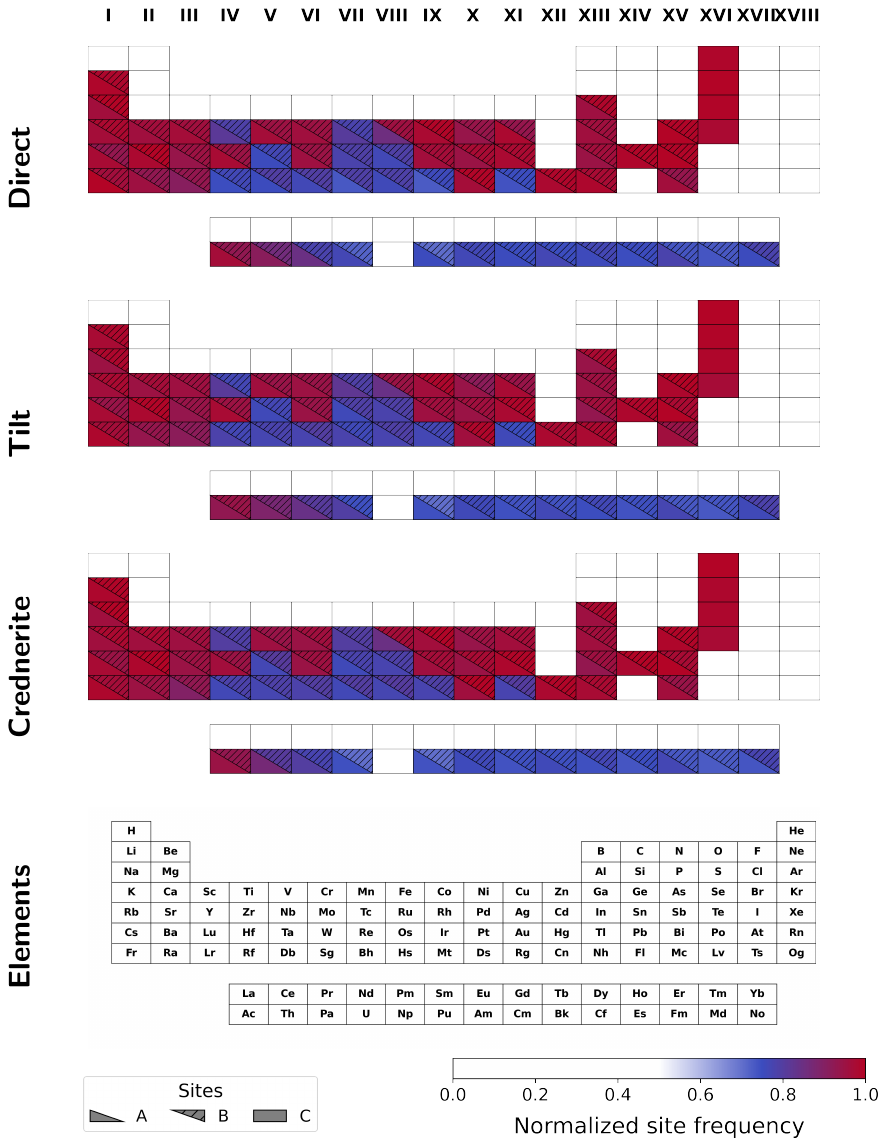}
    \caption{Periodic table heatmaps showing the distribution of elements across the A-, B-, and C-sites in delafossite structures. Rows correspond to stacking types (Direct, Tilt, and Crednerite). Site assignments are indicated by geometric shapes (A-site: lower-left triangle; B-site: upper-right triangle; C-site: full square). Colormap hue (coolwarm) encodes the normalized frequency of occupancy for each site type, with A-, B-, and C-sites normalized separately. Roman numerals above each table denote group numbers. The legend and colorbar provide site definitions and the quantitative scale.}
    \label{fig:periodic_table_structure_stacking_type}
\end{figure}

Each structure in the dataset is assigned a unique identifier in the form: \texttt{A-B-C\_structure\_type}. This identifier is used for reproducibility and dataset management. The distributions of $A$-, $B$-, and $C$-site elements across the three stacking types are summarized in the period table heatmaps (see  Fig.~\ref{fig:periodic_table_structure_stacking_type}). The first three panels show the frequencies of the elements in direct, tilt, and Crednerite structures - a shared color map indicates the frequency with which each element appears, allowing the distribution of site occupations to be compared easily across stacking types. The fourth panel provides an annotated periodic table showing all elements included in the dataset.  For every $A\!-\!B\!-\!C$ atomic triplet, we compute three geometric angles using using the \texttt{pymatgen} Python package~\cite{ong_python_2013}. These angle-based descriptors provide physically interpretable concepts for the Concept Whitening layer and allow latent features to align with meaningful structural quantities. 

This dataset is used to evaluate the effectiveness of Concept Whitening in graph neural networks. We first use it to benchmark CW-GNNs against standard batch-normalized GNNs for stacking-type classification. We then extend the approach to magnetic-ordering prediction, in which structural motifs and composition jointly determine the underlying magnetic ground states. Because layer stacking governs exchange-coupling pathways and spin interactions, an accurate classification of the stacking type is the first necessary step toward modeling the magnetic behavior of delafossite compounds.

%\FloatBarrier
%\subsection{Delafossite Magnetic Ordering Dataset}
%\label{sec:datasets_magnetic_ordering}
We also constructed a second dataset of magnetic delafossites to evaluate the CW-GNN's performance on a more complex, physics-driven classification task. From the full set of relaxed structures, we analyzed 9,498 compounds with triangular lattices using DFT combined with the magnetic force theorem to obtain magnetic solutions. For each structure, four exchange-coupling parameters ($J_1$, $J_2$, $J_3$, and $J_{\mathrm{out}}$) were extracted from the Heisenberg model to quantify in-plane and out-of-plane magnetic interactions. These exchange couplings, together with element-specific $d$-electron features form physically meaningful concepts used in the CW layer for magnetic ordering prediction. Methods section provides additional computational details of the magnetic calculations.

\subsection{GNN Model Performance and Comparison for Structure Classification}
\label{sec:results_and_discussion_benchmarking}

We benchmarked two graph-based models, the baseline batch-normalized GNN (BN-GNN)~\cite{santurkar_how_2018, ioffe_batch_2015} and the CW-GNN, on the full set of 33,520 relaxed delafossite structures labeled by their stacking type: direct, tilt, or Crednerite. Stacking-type prediction provides a controlled baseline task because the three geometries exhibit well-defined and visually distinct connectivity motifs (Fig.~\ref{fig:stacking_graph_grid}).
This makes the task ideal for testing whether the CW module improves stability and interpretability without sacrificing accuracy.

Under identical training conditions, the BN-GNN achieved a validation accuracy of 85.58\%, while the CW-GNN reached 84.94\%. Despite the small difference in accuracy (0.6\%), the CW-GNN demonstrated substantially greater stability across a wide range of perturbations. Quantitative comparisions for all perturbation tests are provided in Supplementary Figures~\ref{fig:SI_benchmarking_concept_weight_val_acc}, \ref{fig:SI_benchmarking_concept_noise_val_acc}, \ref{fig:SI_benchmarking_label_corruption_val_acc}, and \ref{fig:SI_benchmarking_data_fraction_val_acc}. 
%such as noise in the training data. This suggests that it provides more reliable and interpretable structural representations. The specific perturbations and their effects are discussed in detail below. 
%These tests quantify how each architecture responds to variations in hyperparameters, noisy concept information, corrupted labels, and a reduced dataset size.
Supplementary Table~\ref{tab:benhmarking_model_hyperparams} provides all hyperparameters and architectural choices for both models.
To systematically assess the impact of CW into a GNN, we designed four controlled perturbation studies that probe model robustness to typical source of variability in datasets. These sources include: (1) varying the concept-loss weight, (2) adding Gaussian noise to the training concepts, (3) introducing random label corruption, and (4) reducing the size of the available training set. These tests allow us to evaluate whether CW improves stability relative to a standard Batch Normalization-based architecture and whether the CW layer preserves meaningful structural information under perturbations.
%\begin{enumerate}
%    \item concept loss weight,
%    \item Gaussian noise added to training concepts,
%    \item training label corruption, and
%    \item full dataset size.
%\end{enumerate}

For each perturbation setting, we compared the CW-GNN to a BN-GNN baseline in which the CW layer was replaced with Batch Normalization. This ensures that any performance differences arose solely from the concept-whitening mechanism. These benchmarks serve two purposes: to establish whether the CW mechanism provides systematic benefits under controlled perturbations, and to motivate its application to more complex crystalline datasets. In this task, the concepts correspond to three $A$-$B$-$C$ triplet angles (Fig.~\ref{fig:stacking_graph_grid}). Each concept captures one such angle, and the concept metrics (R$^2$) quantifies how strongly any latent dimension of the model correlates with that geometric descriptor. A higher R$^2$ value indicates that the latent space has developed an internal axis that varies predictably with the underlying physical quantity, which demonstrates learned geometric structure. The corresnpoding R$^2$ curves can be found in Supplementary Figures~\ref{fig:SI_benchmarking_concept_weight_concepts}, \ref{fig:SI_benchmarking_concept_noise_concepts}, \ref{fig:SI_benchmarking_label_corruption_concepts}, and \ref{fig:SI_benchmarking_data_fraction_concepts}.

\subsubsection*{Concept Loss Weight}
\label{sec:results_benchmarking_concept_loss_weight}

The total training loss was defined as a weighted sum of the classification 
loss and the concept alignment loss:
\[
\mathcal{L} = \lambda_c \, \mathcal{L}_{\text{concept}} + \lambda_{cls} \, \mathcal{L}_{\text{classification}},
\]
where $\lambda_c$ controls the strength of concept alignment and $\lambda_{cls} = 1.0$ by default. 
We varied $\lambda_c$ over a range of values while keeping all other training 
parameters constant to assess how concept whitening affects this trade-off.
%The final test-set classification accuracy was used as the primary performance metric. 
The CW-GNN maintained consistently high and stable accuracy across the full range of concept-weight values, varying only slightly between about 0.81 and 0.85. This stability indicates that increasing the strength of the concept-alignment loss does not negatively affect the model's predictive performance.
See Supplementary Fig.~\ref{fig:SI_benchmarking_concept_weight_val_acc} for details.

In contrast, the BN-GNN displayed a pronounced sensitivity to $\lambda_c$. At very small values, such as $\lambda_c = 10^{-4}$, its accuracy dropped to $\sim$0.68, which is more than 15 percentage points below the CW-GNN at the same setting. Although performance improved as $\lambda_c$ increased, its accuracy remained more variable overall and matched CW-GNN accuracy only at sufficiently large weights (see Fig.~\ref{fig:SI_benchmarking_concept_weight_val_acc}). 
%The trends summarized here and visualized in Figure \ref{fig:SI_benchmarking_concept_weight_val_acc} highlight the robustness of the CW-GNN compared to the stronger hyperparameter dependence of the BN-GNN.

Furthermore, the concept-latent alignment metrics ($R^2$) highlight the differences between the two architectures. While  the CW-CNN produced stable concept alignment across all $\lambda_c$ values, the BN-GNN showed large fluctuations and inconsistent alignment. This comparison shows that CW not only keeps accuracy but also maintains reliable concept structure even when the alignment weight is changed.

\subsubsection*{Concept Noise}
\label{sec:results_benchmarking_concept_noise}

To evaluate model robustness to imperfect concept information, Gaussian noise was added to the training set concept vectors. Specifically, for each concept vector $\mathbf{c}$, independent noise $\boldsymbol{\epsilon} \sim \mathcal{N}(0, \sigma^2)$ was sampled for each element, producing a perturbed vector
\[
\mathbf{c}' = \mathbf{c} + \boldsymbol{\epsilon}.
\]
The standard deviation $\sigma$ controlled the magnitude of the noise, and only training set concepts were perturbed; validation and test concepts remained unchanged. This procedure allowed us to assess the sensitivity of CW-GNN and BN-GNN models to noisy concept information while keeping all other training parameters fixed.

Using the final test-set classification accuracy as the performance metric, we observed that CW-GNN remained robust under increasing levels of concept noise. Across noise magnitudes from $0.01\sigma$ to $100\sigma$, CW-GNN maintained relatively stable performance, with accuracy decreasing only moderately from approximately 0.85 at low noise to about 0.72 at the highest noise level. This pattern indicates that even substantial perturbations to the concept space do not severely disrupt the predictive behavior of the CW-GNN.

In contrast, the BN-GNN demonstrated a stronger dependence on concept noise. While its accuracy was comparable to CW-GNN at very low noise levels (for example, around 0.84 at $0.01\sigma$ and 0.85 at $0.1\sigma$), its performance deteriorated more quickly as noise increased. At $100\sigma$, BN-GNN accuracy fell to roughly 0.68, which is several percentage points lower than the CW-GNN under the same conditions. These results, summarized here and visualized in Figure \ref{fig:SI_benchmarking_concept_weight_val_acc}, highlight the improved resilience of concept whitening when concepts may be noisy or only partially informative for the downstream prediction task.

The concept $R^2$ metrics showed trends consistent with the accuracy results discussed above. As illustrated in Figure \ref{fig:SI_benchmarking_concept_weight_concepts}. CW-GNN maintained higher concept $R^2$ values under low to moderate noise, indicating that it continued to capture meaningful atomic triplet angle structure even when the concepts were partially corrupted. In contrast, BN-GNN aligned less reliably with the underlying concepts in these regimes. At the highest noise levels, both models experienced substantial reductions in concept learning, demonstrating the practical limits of interpretability when the concept space becomes severely distorted.

\subsubsection*{Label Corruption}
\label{sec:results_benchmarking_label_corruption}

To evaluate model robustness to noisy supervision, a fraction of the training labels was randomly corrupted. For each training sample, the original label $y$ was replaced with a randomly selected alternative class with probability $p$, while all validation and test labels were kept unchanged. The corruption probabilities ranged from $p = 0.01$ to $p = 1.0$, spanning minimal to extreme levels of label noise.

As shown in Figure \ref{fig:SI_benchmarking_label_corruption_val_acc}, CW-GNN and BN-GNN achieved similar performance across low and moderate corruption levels, with both models maintaining high accuracy at small values such as $p = 0.01$. At severe corruption levels, however, both models failed to retain predictive signal, with accuracy dropping to near-random values once $p$ reached $0.75$ and collapsing entirely at $p = 1.0$. These trends indicate that concept whitening does not introduce additional sensitivity to label noise and that both architectures break down only when the supervision signal becomes overwhelmingly unreliable.

\subsubsection*{Dataset Size}
\label{sec:results_benchmarking_dataset_size}

To evaluate the effect of dataset size on model performance, we conducted a split-dataset benchmarking experiment. Subsets of the full dataset were created by randomly sampling fixed fractions of the available data ($1\%, 10\%, 25\%, 50\%, 75\%, 100\%$). For each subset, the relative proportions of the training, validation, and test sets were preserved so that performance differences could be attributed directly to the amount of training data rather than changes in data partitioning. This procedure provided a controlled assessment of how model performance scales with dataset size.

As shown in Figure \ref{fig:SI_benchmarking_data_fraction_val_acc}, both models experienced reduced accuracy at the smallest dataset fraction, with CW-GNN reaching approximately 0.84 and BN-GNN around 0.82 at the $1\%$ subset. Accuracy increased for both models as the available data grew, and CW-GNN consistently achieved higher performance in the medium and large data regimes. For instance, at the $50\%$ subset, CW-GNN attained an accuracy of about 0.84 compared to 0.79 for BN-GNN, and at the $75\%$ subset, CW-GNN again outperformed BN-GNN with accuracies of roughly 0.84 and 0.82, respectively. These results indicate that although both models remain robust across a wide range of dataset sizes, CW-GNN exhibits superior scalability and more reliable gains as additional data become available.

Taken together, these benchmarking results demonstrate that the CW-GNN exhibits stronger overall robustness than the BN-GNN. Across variations in concept-weight strength, injected concept noise, label corruption, and dataset size, the CW-GNN maintained stable predictive performance and showed reduced sensitivity to hyperparameter choices. Establishing this reliability is especially important for delafossite materials, where accurately determining the stacking type is a necessary precursor to understanding their magnetic behavior, since the underlying crystal structure strongly influences the ground-state magnetic ordering. Building on this foundation, we next examine a more complex task by training a deeper CW-GNN designed to capture the richer set of magnetic-ordering concepts associated with these materials.

\subsection{Magnetic Ordering Classification}
\label{sec:results_and_discussion_magnetic_ordering}

The CW-GNN used for magnetic-ordering classification has three hidden layers, and the intermediate dimensions were selected to balance capacity and computational cost. The model employed mean pooling, a dropout rate of 0.2, and Adam with an initial learning rate of $10^{-4}$. All architectural and training hyperparameters are provided in Supplementary Table~\ref{tab:magnetic_hyperparams}.

\begin{figure}[htbp]
    \centering
    \includegraphics[width=\linewidth]{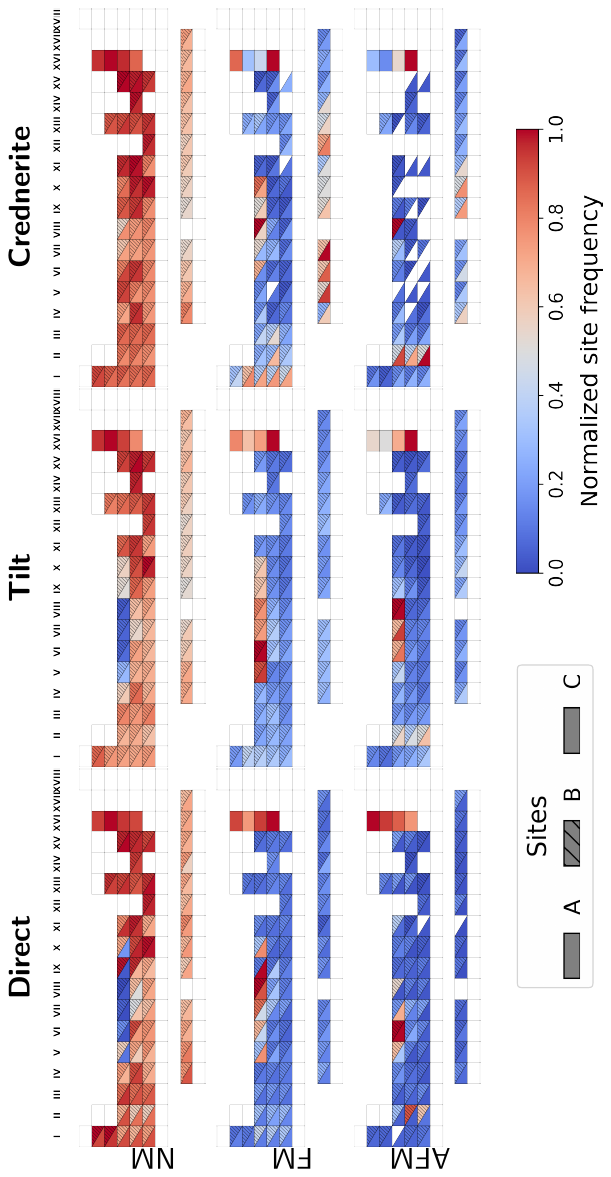}
    \caption{Periodic table heatmaps showing the normalized frequencies of element occupancy across magnetic orderings and stacking types for 33,520 relaxed delafossite structures. Rows correspond to non-magnetic, ferromagnetic, and antiferromagnetic orderings, while columns correspond to Direct, Tilt, and Crednerite structures. Site occupancies are distinguished by plotting conventions: A-site (lower-left triangle), B-site (upper-right triangle), and C-site (full square). Colors follow the coolwarm colormap, with the gradient representing normalized occupancy frequencies, computed separately for A-, B-, and C-sites.}  
    \label{fig:magnetic_ordering_cls_periodic_table}
\end{figure}

The CW-GNN model trained on the magnetic ordering dataset exhibited steady improvement in validation accuracy, reaching a plateau after approximately 50 epochs. Early stopping was triggered at epoch 86, yielding a maximum test set accuracy of 82.6\%. Figure \ref{fig:magnetic_ordering_cls_periodic_table} summarizes these results by visualizing the predicted magnetic ordering across all compositions and stacking types using a 3×3 grid of periodic tables. The columns correspond to the Direct, Tilt, and Crednerite stacking arrangements, while the rows show the distributions of elements for non-magnetic (NM), ferromagnetic (FM), and antiferromagnetic (AFM) classifications. Across all stacking types, NM behavior is the most prevalent, reflecting the dominance of non-magnetic ground states within the dataset. However, clear chemistry–structure trends emerge in the magnetic subclasses. In Direct stacking, Co and Fe occupying the A site most frequently yield FM behavior, while Mn and Fe in the B site also correlate with FM predictions; in contrast, AFM behavior appears most strongly when Cr occupies either the A or B site. A similar pattern is observed in Tilt stacking, where Cr, Mn, and Fe exhibit strong AFM tendencies, whereas V, Cr, and Fe in either cation site more often lead to FM classifications. Crednerite stacking displays the greatest variability, partly due to the smaller number of relaxed structures available, but still shows consistent FM and AFM behavior when Fe is present at the B site. Together, these trends demonstrate that the CW-GNN model achieves strong overall classification performance and reveals clear chemistry–structure relationships across stacking types.

\begin{figure}[htbp]
    \centering
    \includegraphics[width=\linewidth]{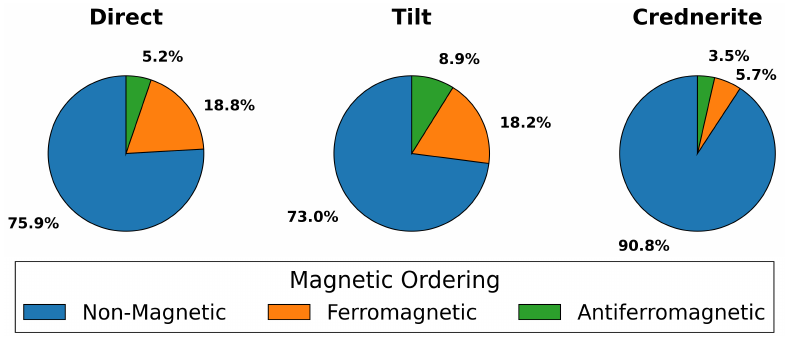}
    \caption{Distribution of magnetic orderings across the three stacking types of delafossites. Each pie chart corresponds to a stacking type (direct, tilt, and crednerite), partitioned into fractions of non-magnetic, ferromagnetic, and antiferromagnetic orderings. The relative wedge sizes represent the observed proportions of each ordering within that stacking type.}
    \label{fig:magnetic_ordering_pies}
\end{figure}

%\textcolor{green}{
The overall distribution of magnetic ordering across the different stacking types is summarized in Figure~\ref{fig:magnetic_ordering_pies}. Non-magnetic (NM) behavior is the dominant ordering in all three stacking types, accounting for 75.9\%, 73.0\%, and 90.8\% of structures with Direct, Tilt, and Crednerite stacking, respectively. Direct and Tilt stackings exhibit comparable fractions of ferromagnetic (FM) structures (18.8\% and 18.2\%), whereas Crednerite stacking shows a markedly lower FM occurrence of 5.7\%. In contrast, antiferromagnetic (AFM) behavior displays a modest divergence between Direct and Tilt stackings (5.2\% and 8.9\%), with Crednerite again exhibiting the lowest fraction at 3.5\%. Collectively, these trends highlight the strong prevalence of NM ordering and indicate that stacking geometry plays a measurable role in shaping FM and AFM behavior.
%}

Additionally, magnetic ordering classification was performed on 142,590 unoptimized delafossite structures, with the full results provided in the Supplementary Information (Figure~\ref{fig:SI_unoptimized_structure_mag_clf}). Notably, all unoptimized Crednerite-type structures were classified as NM, suggesting that the structural distortions present in relaxed geometries are essential for stabilizing FM or AFM behavior. In contrast to the relaxed structures, the choice of chalcogen species played a more significant role in the classifications obtained from the unoptimized dataset.

Element-specific trends were also examined by isolating the contributions of the A- and B-site species. For Direct stacking, Mn and Fe at the B site, as well as Fe and Co at the A site, most frequently yielded FM classifications, while Cr at either site and Mn at the A site were associated with AFM behavior. In Tilt stacking, Fe and Mn at either site were most commonly classified as FM, whereas Cr at either site most frequently appeared in AFM-classified structures.

\begin{figure}[htbp]
    \centering
    \includegraphics[width=\linewidth]{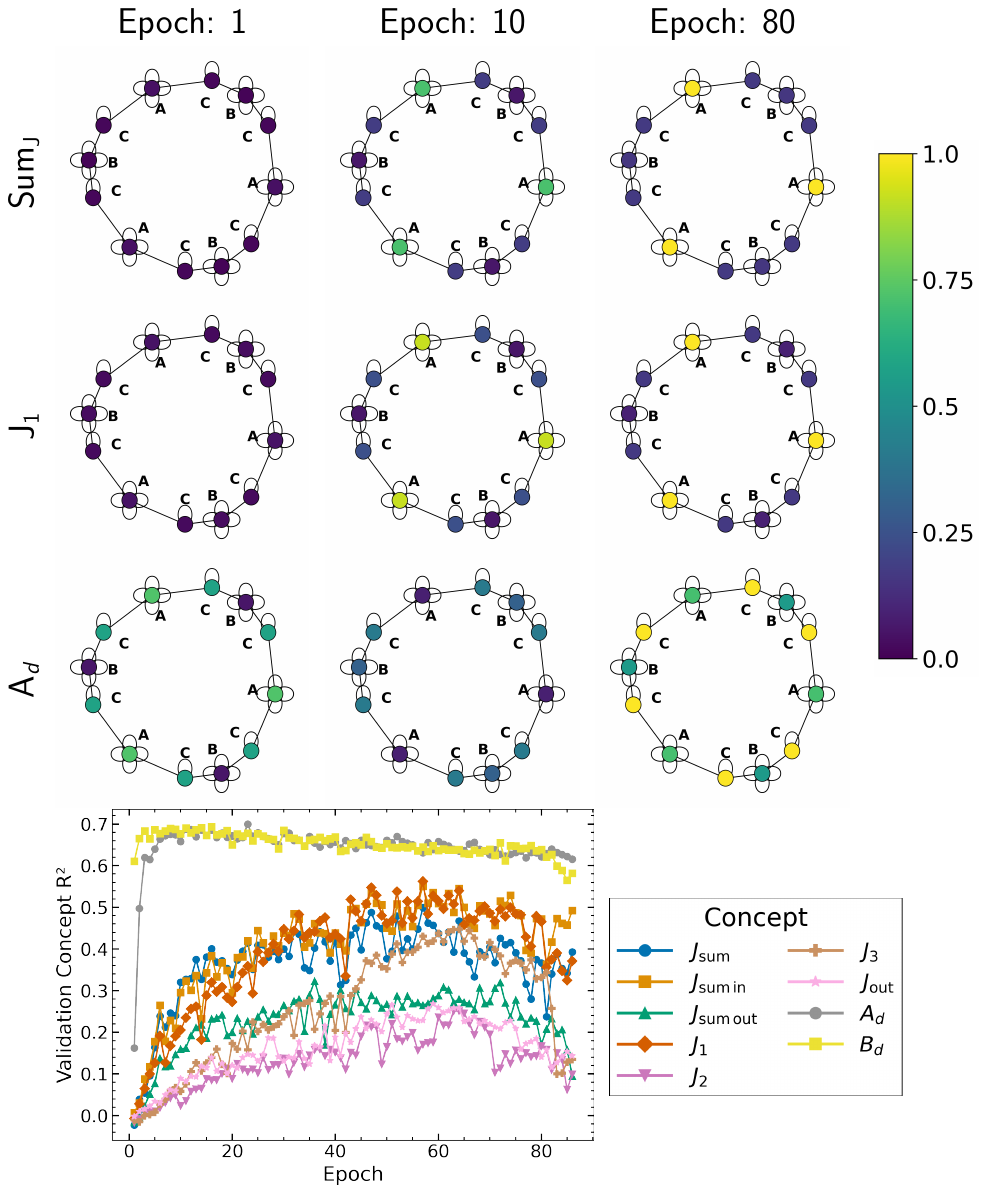}
    \caption{Concept-specific node and edge importance visualizations for the FeCuO$_2$ direct stacking type delafossite's graph representation across training epochs. 
    The top panel shows learned importance for two representative concepts: \texttt{sum\_J} (first row) and \texttt{J1} (second row), at epochs 1, 10, and 80. Node colors represent relative node importance with color scales normalized globally (0--1) for comparability across epochs.  The rightmost column contains the shared colorbar. The bottom section of the figure shows the evolution of the $R^2$ values with each epoch for the nine physical concepts used in training the CW-GNN for magnetic ordering classification. Higher $R^2$ values indicate stronger alignment between the learned latent features and the corresponding physically meaningful concepts.}

    \label{fig:Results_concept_mapping_and_epoch_evolution}
\end{figure}

The nine physically meaningful concepts used in this task exhibited measurable learning, as indicated by their $R^2$ values, though considerable variance was observed across concepts. The strongest alignment occurred for the simpler descriptors, the d-shell valence electrons of the A- and B-site elements, with $R^2 \sim 0.6$. However, signs of overfitting were present, suggesting that these concepts were not consistently mapped onto the graph nodes. As shown in Figure~\ref{fig:Results_concept_mapping_and_epoch_evolution}, the simpler concepts were not appropriately captured in the mapping process, further supporting this observation. In contrast, the interatomic coupling parameters ($J$) demonstrated more stable alignment with the graph representations, despite exhibiting slightly lower $R^2$ values. Among the $J$-related concepts, the in-plane summation $\mathrm{J}_\text{sum in}$, total summation ($J_{\mathrm{sum}}$), and first nearest-neighbor coupling parameter ($J_1$) showed the strongest correspondence, with $R^2$ values ranging from approximately 0.5 to 0.4.

To further illustrate how the CW-GNN develops physically interpretable representations, Figure~\ref{fig:Results_concept_mapping_and_epoch_evolution} also presents the evolution of concept importance, measured by R$^2$ values, across training epochs, together with the physical interactions each concept encodes. These results collectively demonstrate that while simpler electronic descriptors exhibit stronger apparent correlations, the more physically grounded coupling parameters achieve more meaningful and consistent mappings within the learned graph representations.

%\FloatBarrier
In summary, this work demonstrates that concept whitening significantly improves the interpretability of graph neural networks when classifying stacking types and magnetic orderings in delafossite materials. The CW-GNN achieves high accuracy and recovers clear chemistry-structure-magnetism relationships; its concept-aligned latent spaces remain stable during training. Exchange-coupling descriptors exhibit the most coherent alignment, demonstrating that the model internalizes physically meaningful magnetic-interaction pathways. Predictions on unoptimized structures further demonstrate the importance of geometric distortion in stabilizing FM and AFM states.

\FloatBarrier
\section{Methods}
\label{sec:methods}
\subsection{Structure generation}
\label{sec:methods_structure_generation}
Starting from three distinct stacking types of delafossite structures (ABO$_2$) - direct (D), tilt (T), and crednerite (C)- we systematically substituted the A and B sites with 57 different elements: Ag, Al, As, Au, Ba, Bi, Ca, Ce, Co, Cr, Cs, Cu, Dy, Er, Eu, Fe, Ga, Gd, Hf, Hg, Ho, In, Ir, K, La, Li, Lu, Mn, Mo, Na, Nb, Nd, Ni, Os, Pd, Pr, Pt, Rb, Re, Rh, Sb, Sc, Sm, Sn, Sr, Ta, Tb, Tc, Ti, Tl, Tm, V, W, Y, Yb, or Zr. The O site was replaced with O, S, Se, or Te. Structures with identical A and B species (A = B) were excluded, while both ABO$_2$ and BAO$_2$ configurations were considered. This procedure yielded a total of $3 \times 57 \times 56 \times 4 = 38{,}304$ initial structures. All systems were fully relaxed with respect to both atomic positions and cell parameters, and only converged structures were retained for analysis.
All calculations employed identical parameters: an energy cutoff of 500 eV, a reciprocal-space grid defined by a k-spacing of 0.29 Å$^{-1}$, the PBE \cite{perdew_generalized_1996} within the GGA (Generalized Gradient Approximation) exchange–correlation functional, and a convergence criterion of maximum ionic force below $10^{-2}$ eV/Å.

For unoptimized structure generation, the FeCuO$_2$ delafossite in its three stacking types (direct, tilt, and crednerite) was used as the base template. A Python script was employed to systematically substitute the A (Fe), B (Cu), and C (O) sites with a broad candidate pool of elements. For the A and B sites, this pool included alkali metals (Li, Na, K, Rb, Cs, Fr), transition metals (Al, Ga, In, Sn, Ti, Pb, Bi, Nh, Fl, Mc, Lv), metalloids (B, Si, Ge, As, Sb, Te, Po), lanthanides (La, Ce, Pr, Nd, Pm, Sm, Eu, Gd, Tb, Dy, Ho, Er, Tm, Yb, Lu), and actinides (Ac, Th, Pa, U, Np, Pu, Am, Cm, Bk, Cf, Es, Fm, Md, No, Lr). For the C site, the same candidates (O, S, Se) were considered.

\subsection{Graph Construction}
\label{sec:methods_graph_construction}
Each relaxed crystal structure obtained from \textit{ab initio} calculations was converted into a graph representation $G = (V, E)$, where each node $v \in V$ corresponds to an atom and each edge $e \in E$ represents an interatomic connection based on a distance-based cutoff criterion.

The cutoff distances were determined using the radial distribution function (RDF), with element-specific thresholds set to include the first two coordination peaks plus an additional buffer of $0.1$\r{A}. All neighboring atom pairs within these thresholds were connected by an undirected edge.

Each node was featurized with a fixed-length vector derived from a curated elemental dictionary, which was constructed using elemental data from the Periodic Table of Elements.
% \cite{pubchem_periodic_nodate}. 
The feature vector included atomic number, atomic mass, electronegativity, atomic radius, first ionization energy, electron affinity, and the number of valence electrons in the $s$, $p$, $d$, and $f$ orbitals. The number of unpaired electrons was estimated from orbital occupancies and appended as an additional feature. All features were converted to numerical form and cast to \texttt{float32} precision.

Edges were annotated with two attributes: (1) the interatomic distance and (2) the periodic image vector, which captures the cell displacement and direction due to periodic boundary conditions. Both quantities were computed using the \texttt{pymatgen} Python package \cite{ong_python_2013}. To enforce graph symmetry, all edges were explicitly duplicated in both directions. Additionally, self-loops were added to each node, and all graph data were stored using the Deep Graph Library (DGL)\cite{wang_deep_2020}, with node features assigned to \texttt{g.ndata['feat']} and edge attributes to \texttt{g.edata['length']} and \texttt{g.edata['periodicity']}.

\subsection{Magnetic Dataset Generation}
The magnetic delafossite dataset was constructed using a selection of 9,498 structures from the fully relaxed set. These structures exhibit triangular-lattice geometries, which allowed for the use of a Heisenberg-type Hamiltonian to describe their magnetic interactions: 
%The magnetic Hamiltonian are expanded within Heisenberg's exchange coupling models as
\begin{align}
    E=-J_1 \sum_{\braket{i,j}}\mathbf{e}_i\cdot\mathbf{e}_j -J_2 \sum_{\braket{\braket{i,j}}}\mathbf{e}_i\cdot\mathbf{e}_j -J_3 \sum_{\braket{\braket{\braket{i,j}}}}\mathbf{e}_i\cdot\mathbf{e}_j - J_{\mathrm{out}}\sum_{\braket{i,j}_\perp}\mathbf{e}_i\cdot\mathbf{e}_j + \dots,
\end{align}
where $\mathbf{e}_{i,j}$ are the normalized spin moments at atomic sites $i$ and $j$, $\braket{i,j}$, $\braket{\braket{i,j}}$, and $\braket{\braket{\braket{i,j}}}$ represent the first, second, and third nearest neighbors in the triangular plane. $\braket{i,j}_\perp$ is the out-of-plane nearest neighboring coupling. These magnetic couplings $J_1$, $J_2$, $J_3$, and $J_{\mathrm{out}}$ are evaluated based on the DFT combined with the magnetic force theorem \cite{liechtenstein1987local,szilva2023quantitative}. The Green's functions are expressed within the local orbital basis set in the magnetic force theorem calculations. To build the localized atomic Hamiltonian, the wavefunctions of collinear spin configurations are expressed as linear combinations of pseudo-atomic orbitals implemented in OpenMX DFT package \cite{ozaki_variationally_2003}. The Perdew-Burke-Ernzerhof (PBE) \cite{perdew_generalized_1996} functional within the generalized gradient approximation (GGA) scheme was employed for the exchange-correlation method. The k-mesh was sampled as 60 points per 2$\pi$ $\mathrm{\AA}^{-1}$. Magnetic interactions were extracted from the magnetic force theorem calculation as implemented in the Jx code \cite{yoon_jx_2020}. In this study, the positive sign of $J$ indicates the preference of ferromagnetic (FM) ordering and the negative sign is corresponding to the favorable antiferromagnetic (AFM) state. 

\subsection{Concept Whitening Layer}
\label{sec:methods_cw_layer}
Concept Whitening (CW), introduced by Chen et al.\ for interpretable image recognition \cite{chen_concept_2020}, is a normalization technique that decorrelates latent features and aligns them with user-defined, interpretable concept directions. The CW module consists of two primary components:
\begin{enumerate}
    \item \textbf{Whitening:} Transforms latent features to have zero mean and identity covariance, thereby removing correlations.
    \item \textbf{Orthogonal Rotation:} Rotates the whitened features to align specific dimensions with high-level semantic concepts provided by the user.
\end{enumerate}

More formally, let $X \in \mathbb{R}^{n \times d}$ denote a batch of $n$ latent feature vectors, each of dimension $d$, extracted from the penultimate hidden layer of the model. The whitening step transforms $X$ into $Z \in \mathbb{R}^{n \times d}$ such that the transformed features have zero mean and identity covariance:
\[
Z = (X - \mu) W,
\]
where $\mu \in \mathbb{R}^{1 \times d}$ is the mean vector computed across the batch, and $W \in \mathbb{R}^{d \times d}$ is the whitening matrix satisfying $W^\top W = \Sigma^{-1}$, with $\Sigma$ being the covariance matrix of $X$.

To avoid the computational cost and instability of directly computing the matrix inverse square root, the whitening matrix $W$ is instead approximated using the Newton–Schulz iteration. First, the covariance matrix $\Sigma$ is normalized to improve convergence:
\[
\Sigma_N = \frac{\Sigma}{\operatorname{trace}(\Sigma)}.
\]
Starting with $P_0 = I$, where $I \in \mathbb{R}^{d \times d}$ is the identity matrix, the iteration proceeds as:
\[
P_{k+1} = \frac{3}{2} P_k - \frac{1}{2} P_k^3 \Sigma_N, \quad \text{for } k = 0, 1, \dots, T-1,
\]
where $T$ is the total number of iterations. After $T$ steps, the approximate whitening matrix is given by:
\[
W \approx \frac{P_T}{\sqrt{\operatorname{trace}(\Sigma)}}.
\]
This iterative approach allows efficient and differentiable approximation of $\Sigma^{-1/2}$ without requiring explicit eigendecomposition.

Following whitening, an orthogonal rotation is applied to align the whitened features with interpretable concept directions. This is achieved by updating an orthogonal rotation matrix $R \in \mathbb{R}^{d \times d}$ using gradient-based optimization on the Stiefel manifold. The update rule leverages the Cayley transform to preserve orthogonality:
\[
R' = \left(I + \frac{\tau}{2} A\right)^{-1} \left(I - \frac{\tau}{2} A\right) R,
\]
where $A = \nabla \mathcal{L} R^\top - R \nabla \mathcal{L}^\top$ is a skew-symmetric matrix constructed from the gradient $\nabla \mathcal{L}$ of the concept alignment loss with respect to $R$, and $\tau$ is a step size determined via backtracking line search. This formulation ensures that the updated matrix $R'$ remains orthogonal. We adapted the implementation provided by Proietti et al.~\cite{proietti_explainable_2024} from their work on explainable machine learning with CW in drug discovery to support DGL graph data.

\subsection{GNN Model Architecture}
\label{sec:methods_gnn_model_arch}
Our baseline model is an edge-aware graph neural network composed of multiple EGConv mycom{(first time introduced - explain)} \cite{tailor_we_2022} layers. Each EGConv layer updates node embeddings using both node and edge attributes, followed by batch normalization, ReLU activation, and dropout. Node embeddings are aggregated at the graph level using mean pooling. Two fully connected output heads are applied to the pooled embedding: one for material property prediction and one for concept prediction.

%\textcolor{green}{
%The node-wise formulation of the EGConv layer is given by:}
%\[
%\mathbf{x}_i^{\prime} = {\LARGE ||}_{h=1}^H \sum_{\oplus \in
%\mathcal{A}} \sum_{b = 1}^B w_{i, h, \oplus, b} \;
%\underset{j \in \mathcal{N}(i) \cup \{i\}}{\bigoplus}
%\mathbf{W}_b \mathbf{x}_{j}
%\]
%\textcolor{green}{where $\mathbf{W}_b$ denotes the basis weight,$\bigoplus$ denotes an aggregator, and $w$ is the per-%vertex weighting coefficients across different heads, bases and aggregators.}

In BN-GNN models used in Section \ref{sec:results_and_discussion_benchmarking} the BatchNorm1d layer \cite{ioffe_batch_2015} was used to offer comparable results to iterative normalization implemented in the CW layer (see Section \ref{sec:methods_cw_layer}) and ensure isolation CW's effect on results.
%The formulation of the BatchNorm1d layer is given by:
%\[
%y = \frac{x - E[x]}{\sqrt{\text{Var}[x] + \epsilon}} \times \gamma + \beta
%\]
We note that detailed exploration of different GNN architectures and their relative performance is beyond the scope of this study; such investigations have been extensively performed for materials science applications \cite{reiser_graph_2022, yuan_explainable_2022, proietti_explainable_2024, xie_crystal_2018, wu_chemistry-intuitive_2023}.

\section{Data Availability}
\label{sec:data_availability}
The data generated and/or analysed during the current study are available from the corresponding author on request.

\section{Acknowledgement}
\label{sec:acknowledgement}
This work was supported by the U.S. Department of Energy's Office of Science, Office of Basic Energy Sciences, Materials Sciences and Engineering Division and the Oak Ridge Leadership Computing Facility at the Oak Ridge National Laboratory, which is supported by the Office of Science of the U.S. Department of Energy under Contract No. DE-AC05-00OR22725.

%This research was supported in part by an appointment to the Education Collaboration at ORNL, sponsored by the U.S. Department of Energy and administered by ORISE.
%This research used resources of the Compute and Data Environment for Science (CADES) at the Oak Ridge National Laboratory, which is supported by the Office of Science of the U.S. Department of Energy under Contract No. DE-AC05-00OR22725.

\bibliographystyle{apsrev4-1}
\bibliography{biblio}

@article{liechtenstein1987local,
  title={Local spin density functional approach to the theory of exchange interactions in ferromagnetic metals and alloys},
  author={Liechtenstein, A Il and Katsnelson, MI and Antropov, VP and Gubanov, VA},
    journal = {Journal of Magnetism and Magnetic Materials},
  volume={67},
  number={1},
  pages={65--74},
  year={1987},
}

@article{szilva2023quantitative,
  title={Quantitative theory of magnetic interactions in solids},
  author={Szilva, Attila and Kvashnin, Yaroslav and Stepanov, Evgeny A and Nordstr{\"o}m, Lars and Eriksson, Olle and Lichtenstein, Alexander I and Katsnelson, Mikhail I},
    journal = {Rev. Mod. Phys.},
  volume={95},
  number={3},
  pages={035004},
  year={2023},
}

@article{zhong_explainable_2022,
	title = {Explainable machine learning in materials science},
	volume = {8},
	copyright = {2022 The Author(s)},
	language = {en},
	number = {1},
	urldate = {2025-06-27},
	  journal = {npj Comput. Mater.},
	author = {Zhong, Xiaoting and Gallagher, Brian and Liu, Shusen and Kailkhura, Bhavya and Hiszpanski, Anna and Han, T. Yong-Jin},
	month = sep,
	year = {2022},
	pages = {204},
}

@article{liao_predicting_2022,
	title = {Predicting magnetic anisotropy energies using site-specific spin-orbit coupling energies and machine learning: {Application} to iron-cobalt nitrides},
	volume = {6},
	shorttitle = {Predicting magnetic anisotropy energies using site-specific spin-orbit coupling energies and machine learning},
	number = {2},
	urldate = {2025-06-27},
	  journal = {Phys. Rev. Mater.},
	author = {Liao, Timothy and Xia, Weiyi and Sakurai, Masahiro and Wang, Renhai and Zhang, Chao and Sun, Huaijun and Ho, Kai-Ming and Wang, Cai-Zhuang and Chelikowsky, James R.},
	month = feb,
	year = {2022},
	pages = {024402},
}

@article{butler_machine_2018,
	title = {Machine learning for molecular and materials science},
	volume = {559},
	copyright = {2018 Macmillan Publishers Ltd., part of Springer Nature},
	language = {en},
	number = {7715},
	urldate = {2025-06-30},
	  journal = {Nature},
	author = {Butler, Keith T. and Davies, Daniel W. and Cartwright, Hugh and Isayev, Olexandr and Walsh, Aron},
	month = jul,
	year = {2018},
	pages = {547--555},
}

@article{xie_crystal_2018,
	title = {Crystal {Graph} {Convolutional} {Neural} {Networks} for an {Accurate} and {Interpretable} {Prediction} of {Material} {Properties}},
	volume = {120},
	number = {14},
	urldate = {2025-06-30},
	  journal = {Phys. Rev. Lett.},
	author = {Xie, Tian and Grossman, Jeffrey C.},
	month = apr,
	year = {2018},
	pages = {145301},
}

@article{reiser_graph_2022,
	title = {Graph neural networks for materials science and chemistry},
	volume = {3},
	copyright = {2022 The Author(s)},
	language = {en},
	number = {1},
	urldate = {2025-07-03},
	  journal = {Commun. Mater.},
	author = {Reiser, Patrick and Neubert, Marlen and Eberhard, André and Torresi, Luca and Zhou, Chen and Shao, Chen and Metni, Houssam and van Hoesel, Clint and Schopmans, Henrik and Sommer, Timo and Friederich, Pascal},
	month = nov,
	year = {2022},
	pages = {93},
}

@article{yuan_explainable_2022,
	title = {Explainable graph neural networks for organic cages},
	volume = {1},
	language = {en},
	number = {2},
	urldate = {2025-07-08},
	  journal = {Digit. Discov.},
	author = {Yuan, Qi and Szczypiński, Filip T. and Jelfs, Kim E.},
	month = apr,
	year = {2022},
	pages = {127--138},
}

@article{wu_chemistry-intuitive_2023,
	title = {Chemistry-intuitive explanation of graph neural networks for molecular property prediction with substructure masking},
	volume = {14},
	copyright = {2023 The Author(s)},
	language = {en},
	number = {1},
	urldate = {2025-07-08},
	  journal = {Nat. Commun.},
	author = {Wu, Zhenxing and Wang, Jike and Du, Hongyan and Jiang, Dejun and Kang, Yu and Li, Dan and Pan, Peichen and Deng, Yafeng and Cao, Dongsheng and Hsieh, Chang-Yu and Hou, Tingjun},
	month = may,
	year = {2023},
	pages = {2585},
}

@article{xiao_graph_2024,
	title = {Graph isomorphism network for materials property prediction along with explainability analysis},
	volume = {233},
	urldate = {2025-07-09},
	  journal = {Comput. Mater. Sci.},
	author = {Xiao, Jianping and Yang, Li and Wang, Shuqun},
	month = jan,
	year = {2024},
	pages = {112619},
}

@article{marquardt_crystal_2006,
	series = {Proceedings of the {Fourth} {International} {Symposium} on {Transparent} {Oxide} {Thin} {Films} for {Electronics} and {Optics} ({TOEO}-4)},
	title = {Crystal chemistry and electrical properties of the delafossite structure},
	volume = {496},
	number = {1},
	urldate = {2025-07-16},
	  journal = {Thin Solid Films},
	author = {Marquardt, Meagen A. and Ashmore, Nathan A. and Cann, David P.},
	month = feb,
	year = {2006},
	pages = {146--156},
}

@article{taddee_characterization_2016,
	series = {Proceedings for {International} {Conference} on {Surfaces}, {Coatings} and {Nanostructured} {Materials} ({NANOSMAT}-10, {Manchester}, {UK})},
	title = {Characterization of transparent superconductivity {Fe}-doped {CuCrO2} delafossite oxide},
	volume = {380},
	urldate = {2025-07-17},
	  journal = {Appl. Surf. Sci.},
	author = {Taddee, Chutirat and Kamwanna, Teerasak and Amornkitbamrung, Vittaya},
	month = sep,
	year = {2016},
	pages = {237--242},
}

@article{kresse_ab_1993,
	title = {Ab initio molecular dynamics for liquid metals},
	volume = {47},
	number = {1},
	urldate = {2025-07-17},
	  journal = {Phys. Rev. B},
	author = {Kresse, G. and Hafner, J.},
	month = jan,
	year = {1993},
	pages = {558--561},
}

@article{kresse_ab_1994,
	title = {Ab initio molecular-dynamics simulation of the liquid-metal--amorphous-semiconductor transition in germanium},
	volume = {49},
	number = {20},
	urldate = {2025-07-17},
	  journal = {Phys. Rev. B},
	author = {Kresse, G. and Hafner, J.},
	month = may,
	year = {1994},
	pages = {14251--14269},
}

@article{yoon_jx_2020,
	title = {Jx: {An} open-source software for calculating magnetic interactions based on magnetic force theory},
	volume = {247},
	shorttitle = {Jx},
	urldate = {2025-07-17},
	  journal = {Comput. Phys. Commun.},
	author = {Yoon, Hongkee and Kim, Taek Jung and Sim, Jae-Hoon and Han, Myung Joon},
	month = feb,
	year = {2020},
	pages = {106927},
}

@article{chen_concept_2020,
	title = {Concept whitening for interpretable image recognition},
	volume = {2},
	copyright = {2020 The Author(s), under exclusive licence to Springer Nature Limited},
	language = {en},
	number = {12},
	urldate = {2025-07-22},
	  journal = {Nat. Mach. Intell.},
	author = {Chen, Zhi and Bei, Yijie and Rudin, Cynthia},
	month = dec,
	year = {2020},
	pages = {772--782},
}

@article{proietti_explainable_2024,
	title = {Explainable {AI} in drug discovery: self-interpretable graph neural network for molecular property prediction using concept whitening},
	volume = {113},
	shorttitle = {Explainable {AI} in drug discovery},
	language = {en},
	number = {4},
	urldate = {2025-07-24},
	  journal = {Machine Learning},
	author = {Proietti, Michela and Ragno, Alessio and Rosa, Biagio La and Ragno, Rino and Capobianco, Roberto},
	month = apr,
	year = {2024},
	pages = {2013--2044},
}

@misc{wang_deep_2020,
	title = {Deep {Graph} {Library}: {A} {Graph}-{Centric}, {Highly}-{Performant} {Package} for {Graph} {Neural} {Networks}},
	shorttitle = {Deep {Graph} {Library}},
	urldate = {2025-08-06},
	author = {Wang, Minjie and Zheng, Da and Ye, Zihao and Gan, Quan and Li, Mufei and Song, Xiang and Zhou, Jinjing and Ma, Chao and Yu, Lingfan and Gai, Yu and Xiao, Tianjun and He, Tong and Karypis, George and Li, Jinyang and Zhang, Zheng},
	month = aug,
	year = {2020},
	note = {arXiv:1909.01315},
}

@article{ong_python_2013,
	title = {Python {Materials} {Genomics} (pymatgen): {A} robust, open-source python library for materials analysis},
	volume = {68},
	shorttitle = {Python {Materials} {Genomics} (pymatgen)},
	urldate = {2025-08-06},
	  journal = {Comput. Mater. Sci.},
	author = {Ong, Shyue Ping and Richards, William Davidson and Jain, Anubhav and Hautier, Geoffroy and Kocher, Michael and Cholia, Shreyas and Gunter, Dan and Chevrier, Vincent L. and Persson, Kristin A. and Ceder, Gerbrand},
	month = feb,
	year = {2013},
	pages = {314--319},
}

@article{banik_evaluating_2024,
	title = {Evaluating generalized feature importance via performance assessment of machine learning models for predicting elastic properties of materials},
	volume = {236},
	urldate = {2025-08-17},
	  journal = {Comput. Mater. Sci.},
	author = {Banik, Suvo and Balasubramanian, Karthik and Manna, Sukriti and Derrible, Sybil and Sankaranarayananan, Subramanian K. R. S.},
	month = mar,
	year = {2024},
	pages = {112847},
}

@article{sarada_devi_graph_2024,
	title = {Graph neural network-based multiscale thermal modeling for heterogeneous materials with complex structures},
	volume = {55},
	urldate = {2025-08-17},
	  journal = {Therm. Sci. Eng. Prog.},
	author = {Sarada Devi, C. H. and Sahaaya Arul Mary, S. A. and Karthikeyan, N. and Varalakshmi, S. and Talasila, Vamsidhar and Rama Naidu, G.},
	month = oct,
	year = {2024},
	pages = {102983},
}

@article{gao_molecular_2024,
	title = {Molecular descriptor-enhanced graph neural network for energetic molecular property prediction},
	volume = {67},
	language = {English},
	number = {4},
	urldate = {2025-08-17},
	  journal = {Sci. China Mater.},
	author = {Gao, Tianyu and Ji, Yujin and Liu, Cheng and Li, Youyong},
	month = apr,
	year = {2024},
	pages = {1243--1252},
}

@article{rasool_optimizing_2024,
	title = {Optimizing {GNN} {Architectures} {Through} {Nonlinear} {Activation} {Functions} for {Potent} {Molecular} {Property} {Prediction}},
	volume = {12},
	copyright = {http://creativecommons.org/licenses/by/3.0/},
	language = {en},
	number = {11},
	urldate = {2025-08-17},
	  journal = {Computation},
	author = {Rasool, Areen and Rahman, Jamshaid Ul and Iqbal, Quaid},
	month = nov,
	year = {2024},
	pages = {212},
}

@article{fan_accelerate_2024,
	title = {Accelerate the design of new superhard carbon allotropes in \textit{{Pca}}21 space group: {High}-throughput screening and machine learning strategies},
	volume = {143},
	shorttitle = {Accelerate the design of new superhard carbon allotropes in \textit{{Pca}}21 space group},
	urldate = {2025-08-17},
	  journal = {Diam. Relat. Mater.},
	author = {Fan, Qingyang and Min, Gege and Liu, Li and Zhao, Yingbo and Yu, Xinhai and Yun, Sining},
	month = mar,
	year = {2024},
	pages = {110928},
}

@article{liu_machine_2023,
	title = {Machine {Learning} for {Perovskite} {Solar} {Cells} and {Component} {Materials}: {Key} {Technologies} and {Prospects}},
	volume = {33},
	shorttitle = {Machine {Learning} for {Perovskite} {Solar} {Cells} and {Component} {Materials}},
	language = {English},
	number = {17},
	urldate = {2025-08-17},
	  journal = {Adv. Funct. Mater.},
	author = {Liu, Yiming and Tan, Xinyu and Liang, Jie and Han, Hongwei and Xiang, Peng and Yan, Wensheng},
	month = apr,
	year = {2023},
}

@article{yang_machine_2023,
	title = {Machine learning prediction of specific capacitance in biomass derived carbon materials: {Effects} of activation and biochar characteristics},
	volume = {331},
	shorttitle = {Machine learning prediction of specific capacitance in biomass derived carbon materials},
	language = {English},
	urldate = {2025-08-17},
	  journal = {Fuel},
	author = {Yang, Xuping and Yuan, Chuan and He, Sirong and Jiang, Ding and Cao, Bin and Wang, Shuang},
	month = jan,
	year = {2023},
	pages = {125718},
}

@article{nevolianis_multi-fidelity_2025,
	title = {Multi-fidelity graph neural networks for predicting toluene/water partition coefficients},
	volume = {17},
	number = {1},
	urldate = {2025-08-17},
	  journal = {Journal of Cheminformatics},
	author = {Nevolianis, Thomas and Rittig, Jan G. and Mitsos, Alexander and Leonhard, Kai},
	month = aug,
	year = {2025},
	pages = {123},
}

@article{yuan_explainability_2023,
	title = {Explainability in {Graph} {Neural} {Networks}: {A} {Taxonomic} {Survey}},
	volume = {45},
	shorttitle = {Explainability in {Graph} {Neural} {Networks}},
	language = {English},
	number = {5},
	urldate = {2025-08-18},
	  journal = {IEEE Trans. Pattern Anal. Mach. Intell.},
	author = {Yuan, Hao and Yu, Haiyang and Gui, Shurui and Ji, Shuiwang},
	month = may,
	year = {2023},
	pages = {5782--5799},
}

@article{noh_anisotropic_2009,
	title = {Anisotropic {Electric} {Conductivity} of {Delafossite} {PdCoO2} {Studied} by {Angle}-{Resolved} {Photoemission} {Spectroscopy}},
	volume = {102},
	language = {English},
	number = {25},
	urldate = {2025-08-18},
	  journal = {Phys. Rev. Lett.},
	author = {Noh, Han-Jin and Jeong, Jinwon and Jeong, Jinhwan and Cho, En-Jin and Kim, Sung Baek and Kim, Kyoo and Min, B. I. and Kim, Hyeong-Do},
	month = jun,
	year = {2009},
	pages = {256404},
}

@article{lunca-popa_tuneable_2020,
	title = {Tuneable interplay between atomistic defects morphology and electrical properties of transparent p-type highly conductive off-stoichiometric {Cu}-{Cr}-{O} delafossite thin films},
	volume = {10},
	language = {English},
	number = {1},
	urldate = {2025-08-18},
	  journal = {Sci. Rep.},
	author = {Lunca-Popa, Petru and Botsoa, Jacques and Bahri, Mounib and Crepelliere, Jonathan and Desgardin, Pierre and Audinot, Jean-Nicolas and Wirtz, Tom and Arl, Didier and Ersen, Ovidiu and Barthe, Marie-France and Lenoble, Damien},
	month = jan,
	year = {2020},
	pages = {1416},
}

@article{matasov_3r-cucro2_2024,
	title = {{3R}-{CuCrO2} delafosite: crystal growth, crystal structure, dielectric and {DC} conductivity properties},
	volume = {4},
	shorttitle = {{3R}-{CuCrO2} delafosite},
	language = {en},
	number = {1},
	urldate = {2025-08-18},
	  journal = {Discov. Mater.},
	author = {Matasov, Anton and Bush, Alexander and Kozlov, Vladislav and Stash, Adam},
	month = nov,
	year = {2024},
	pages = {73},
}

@article{el-bassuony_influence_2022,
	title = {Influence of {Elastic} and {Optical} {Properties} on {AgFeO2} and {AgCrO2} {Delafossite} to be {Applied} in {High}-{Frequency} {Applications}},
	volume = {74},
	language = {English},
	number = {7},
	urldate = {2025-08-18},
	  journal = {JOM},
	author = {El-Bassuony, Asmaa A. H. and Abdelsalam, H. K. and Gamal, W. M.},
	month = jul,
	year = {2022},
	pages = {2656--2664},
}

@article{gharbi_exploring_2025,
	title = {Exploring delafossite oxide {Li2Fe2O4}: {Structural}, surface morphology, and optical characteristics obtained through auto-combustion technique},
	volume = {174},
	shorttitle = {Exploring delafossite oxide {Li2Fe2O4}},
	language = {English},
	urldate = {2025-08-18},
	  journal = {Inorg. Chem. Commun.},
	author = {Gharbi, Souha and Messoudi, J. and Dhahri, R. and Gharbi, S. and Taboukhat, S. and Dhahri, E. and Barille, R. and Sahraoui, B.},
	month = apr,
	year = {2025},
	pages = {114022},
}

@article{ketfi_insight_2023,
	title = {Insight into the spin-polarized structural, optoelectronic, magnetic, thermodynamic, and thermoelectric properties of {PdBO2} ({B} = {Al}, {Cr}, and {Rh}) Delafossite semiconductor},
	volume = {55},
	language = {en},
	number = {11},
	urldate = {2025-08-18},
	  journal = {Opt. Quantum Electron.},
	author = {Ketfi, Mohammed Elamin and Essaoud, Saber Saad and Al Azar, Said and Al-Reyahi, Anas Y. and Mousa, Ahmad A. and Mufleh, Ahmad},
	month = sep,
	year = {2023},
	pages = {1013},
}

@article{fujii_tlybse2_2025,
	title = {{TlYbSe2} as a member of the {J}=12 triangular-lattice {Yb} delafossite family: {From} spin liquid to field-induced magnetic order},
	volume = {112},
	shorttitle = {{TlYbSe2} as a member of the {J}=12 triangular-lattice {Yb} delafossite family},
	language = {English},
	number = {2},
	urldate = {2025-08-18},
	  journal = {Phys. Rev. B},
	author = {Fujii, T. and Pillaca, M. and Baertl, F. and Sichelschmidt, J. and Luther, S. and Rosner, H. and Strydom, A. M. and Wosnitza, J. and Kuehne, H. and Doert, Th and Baenitz, M.},
	month = jul,
	year = {2025},
	pages = {024426},
}

@article{bouda_unexpected_2019,
	title = {Unexpected magnetic behavior of {Ga} doped {CuFe1}-{xGaxO2} delafossite, x=0.04: {First} principle calculation and {Monte} {Carlo} simulation},
	volume = {134},
	shorttitle = {Unexpected magnetic behavior of {Ga} doped {CuFe1}-{xGaxO2} delafossite, x=0.04},
	language = {English},
	number = {10},
	urldate = {2025-08-18},
	  journal = {Eur. Phys. J. Plus},
	author = {Bouda, H. and Bahlagui, T. and Masrour, R. and Bahmad, L. and El Kenz, A. and Benyoussef, A.},
	month = oct,
	year = {2019},
	pages = {543},
}

@article{benreguia_optical_2024,
	title = {Optical and electrical properties of the delafossite {CuCrO2} synthesized by co-precipitation},
	volume = {162},
	urldate = {2025-08-18},
	  journal = {Inorg. Chem. Commun.},
	author = {Benreguia, N. and Rekhila, G. and Younes, A. and Abdi, A. and Trari, M.},
	month = apr,
	year = {2024},
	pages = {112154},
}

@article{ozaki_variationally_2003,
	title = {Variationally optimized atomic orbitals for large-scale electronic structures},
	volume = {67},
	number = {15},
	urldate = {2025-08-21},
	  journal = {Phys. Rev. B},
	author = {Ozaki, T.},
	month = apr,
	year = {2003},
	pages = {155108},
}

@article{perdew_generalized_1996,
	title = {Generalized {Gradient} {Approximation} {Made} {Simple}},
	volume = {77},
	number = {18},
	urldate = {2025-08-21},
	  journal = {Phys. Rev. Lett.},
	author = {Perdew, John P. and Burke, Kieron and Ernzerhof, Matthias},
	month = oct,
	year = {1996},
	pages = {3865--3868},
}

@inproceedings{santurkar_how_2018,
	title = {How Does Batch Normalization Help Optimization?},
	volume = {31},
	urldate = {2025-11-24},
	booktitle = {Advances in {Neural} {Information} {Processing} {Systems}},
	author = {Santurkar, Shibani and Tsipras, Dimitris and Ilyas, Andrew and Madry, Aleksander},
	year = {2018},
}

@misc{tailor_we_2022,
	title = {Do {We} {Need} {Anisotropic} {Graph} {Neural} {Networks}?},
	urldate = {2025-11-24},
	author = {Tailor, Shyam A. and Opolka, Felix L. and Liò, Pietro and Lane, Nicholas D.},
	month = may,
	year = {2022},
	note = {arXiv:2104.01481},
}

@misc{ioffe_batch_2015,
	title = {Batch {Normalization}: {Accelerating} {Deep} {Network} {Training} by {Reducing} {Internal} {Covariate} {Shift}},
	shorttitle = {Batch {Normalization}},
	doi = {10.48550/arXiv.1502.03167},
	urldate = {2025-11-25},
	author = {Ioffe, Sergey and Szegedy, Christian},
	month = mar,
	year = {2015},
	note = {arXiv:1502.03167},
}

\clearpage

%\end{document}

%\section*{\large Author Contributions}

\section{Figures}

\begin{figure}[htbp]
    \centering
    \includegraphics[width=\linewidth]{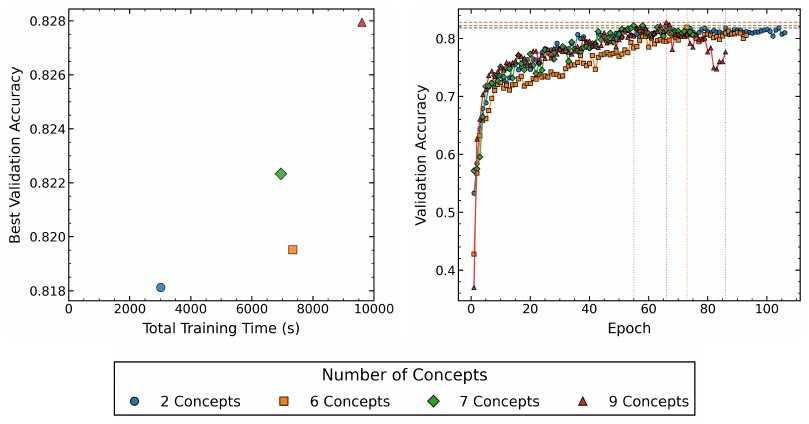}
    \caption{Accuracy--performance trade-off as a function of the number of concepts used during model training. Four distinct concept sets were evaluated: (1) \textit{d}-shell valence electrons of A- and B-site elements, (2) all valence electrons across each shell of A- and B-site elements, (3) exchange coupling parameters ($J$) for first, second, and third nearest neighbors---considering in-plane, out-of-plane, and total sums---and (4) a combined set including both coupling parameters and \textit{d}-shell valence electron concepts. The left panel shows the best validation accuracy achieved as a function of total training time, while the right panel illustrates the evolution of validation accuracy with epochs, with horizontal lines indicating the best validation accuracy for each concept set and vertical lines indicating the epoch at which the best validation accuracy was obtained.}
    
    \label{fig:Results_concept_number_accuracy_performance_trade_off}
\end{figure}

\clearpage
\onecolumngrid

\section*{Supplementary Information}

\setcounter{figure}{0}
\setcounter{table}{0}
\renewcommand{\thefigure}{S\arabic{figure}}
\renewcommand{\thetable}{S\arabic{table}}

\subsection{Magnetic-Ordering Predictions Based on Unoptimized Delafossite Structures}

We performed the magnetic ordering classification on unoptimized delafossite structures for two main reasons: (i) to demonstrate that CW-GNN can be applied prior to full structural relaxation, which greatly reduces computational cost and enables large-scale materials screening, and (ii) to highlight the role of geometric distortions in determining the magnetic ordering, in addition to the influence of elemental species at the $A$, $B$, and $C$ sites. Furthermore, because of the procedure described in Sections~\ref{sec:methods_structure_generation} and \ref{sec:methods_graph_construction}, the resulting graphs for unoptimized structures have identical connectivity, differing only in node attributes. This allows for consistent comparisons between relaxed and unrelaxed datasets.

\begin{figure}[htbp]
    \centering
    \includegraphics[width=\linewidth]{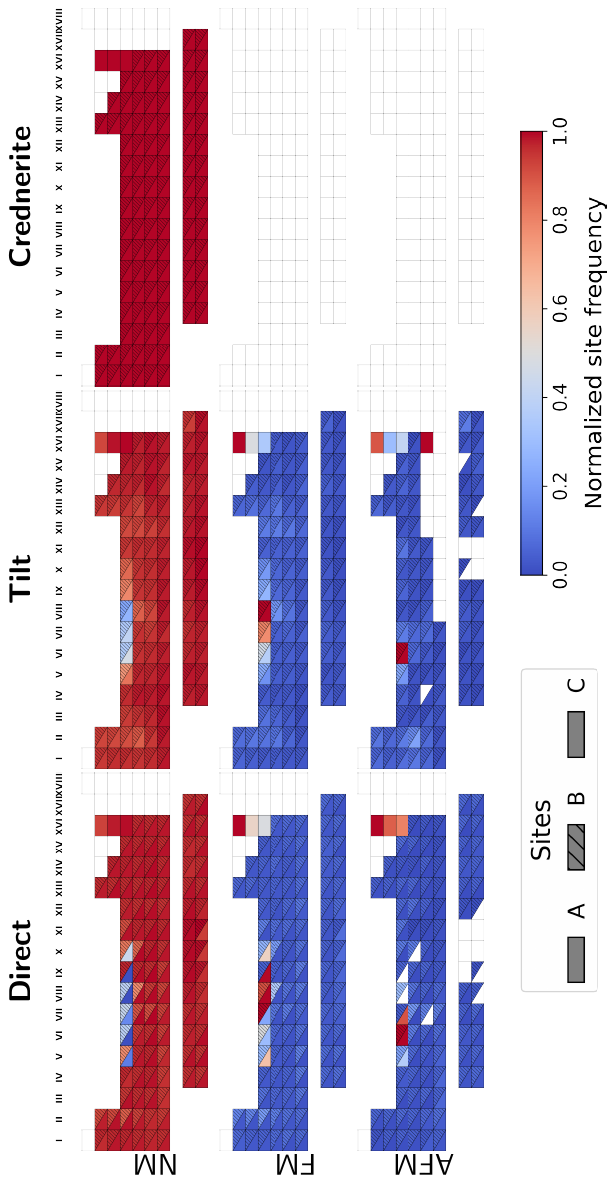}
    \caption{Periodic table heatmaps showing the normalized frequencies of element occupancy across magnetic orderings and stacking types for 142,590 unoptimized delafossite structures. Rows correspond to non-magnetic, ferromagnetic, and antiferromagnetic orderings, while columns correspond to Direct, Tilt, and Crednerite structures. Site occupancies are distinguished by plotting conventions: $A$-site (lower-left triangle), $B$-site (upper-right triangle), and $C$-site (full square). Colors follow the coolwarm colormap, with the gradient representing normalized occupancy frequencies, computed separately for $A$-, $B$-, and $C$-sites.}
    \label{fig:SI_unoptimized_structure_mag_clf}
\end{figure}

Figure~\ref{fig:SI_unoptimized_structure_mag_clf} summarizes the predicted magnetic ground states - nonmagnetic (NM), ferromagnetic (FM), and antiferromagnetic (AFM) - for the 142,590 unoptimized delafossite structures across the direct, tilt, and Crednerite stacking types. Almost all crednerite structures are predicted to be nonmagnetic. In constrast to crednerite, the direct and tilt sturctures display substrantial fraction of FM and AFM states. The presence of S, Se, or Te on the C site alters the electronic environment and exchange couplings, leading to noticeably different magnetic distributions relative to O-based compounds. The model identifies expected trends, such as increased magnetic propensity in late transition-metal compositions and suppressed magnetism in closed-shell systems. However, the differences in the magnetic data from the optimized structures are obvious, highlighting the importance of structure optimization.

\begin{table}[htbp]
    \centering
    \caption{CW-GNN and BN-GNN model architecture and training hyperparameters.}
    \label{tab:benhmarking_model_hyperparams}
    \begin{tabular}{l l}
        \hline
        \textbf{Hyperparameter} & \textbf{Value} \\
        \hline
        Number of hidden layers & 2 \\
        Hidden layer dimensions & 128, 128 \\
        Attention heads per layer & 1 \\
        Node pooling & Mean \\
        Dropout & 0.2 \\
        CW layer (\texttt{use\_cw}) & True / False \\
        Optimizer & Adam \\
        Learning rate & $1\times 10^{-4}$ \\
        Batch size & 32 \\
        Number of epochs & 50 \\
        Early stopping patience & 5 \\
        Learning rate decay & Not used \\
        Training/validation/test split & 70\% / 15\% / 15\% \\
        Concept scaling & StandardScaler fitted on training set \\
        \hline
    \end{tabular}
\end{table}

%\begin{sidewaysfigure}[htbp]
%\end{sidewaysfigure}

\subsection{Benchmarking of Model Robustness}

\begin{figure}
    \centering
    \includegraphics[width=0.8\linewidth]{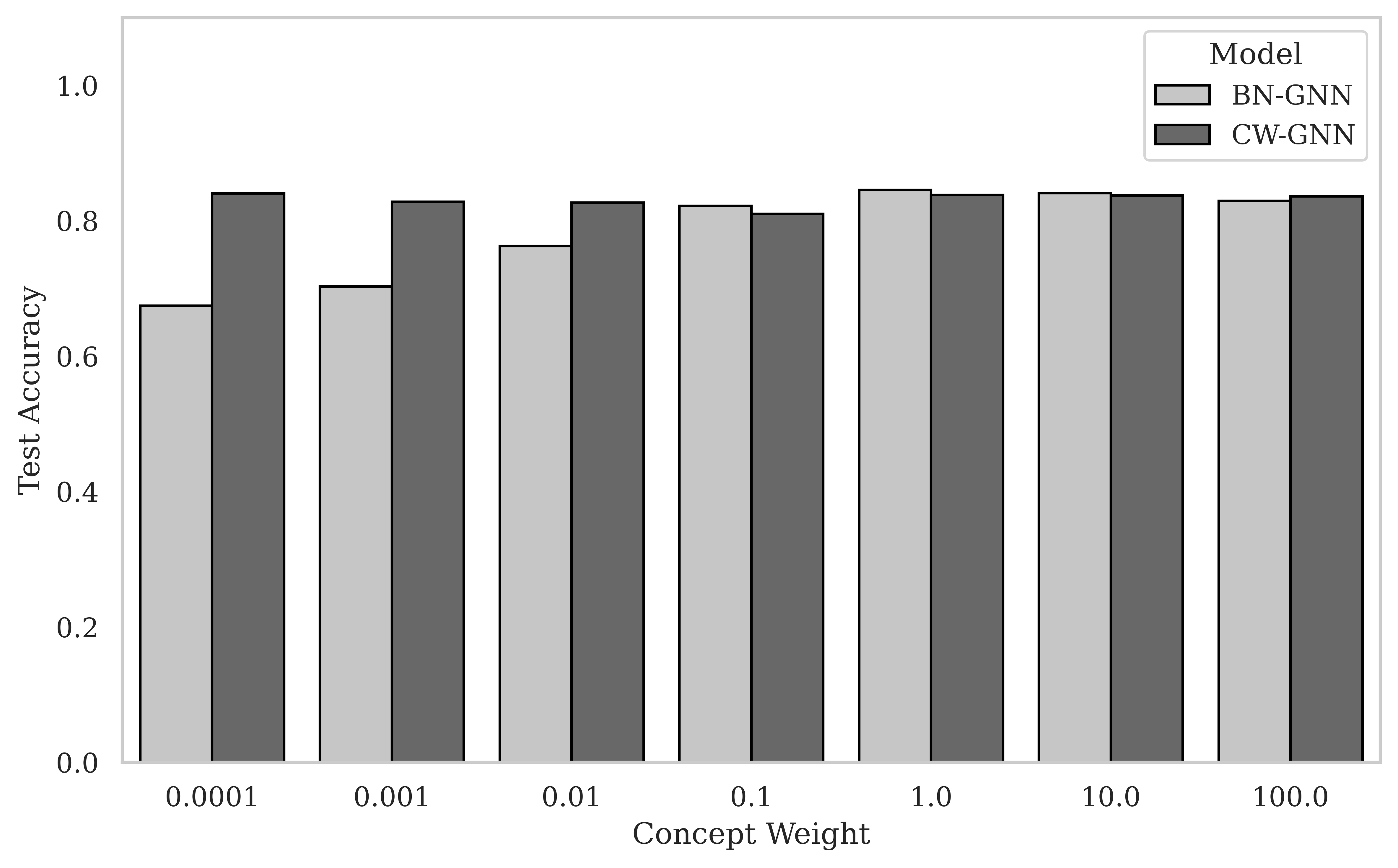}
    \caption{Bar plot of final validation accuracy for BN-GNN and CW-GNN models across values of the concept loss weight parameter ($\lambda_c$).}
    \label{fig:SI_benchmarking_concept_weight_val_acc}
\end{figure}

\begin{figure}
    \centering
    \includegraphics[width=0.8\linewidth]{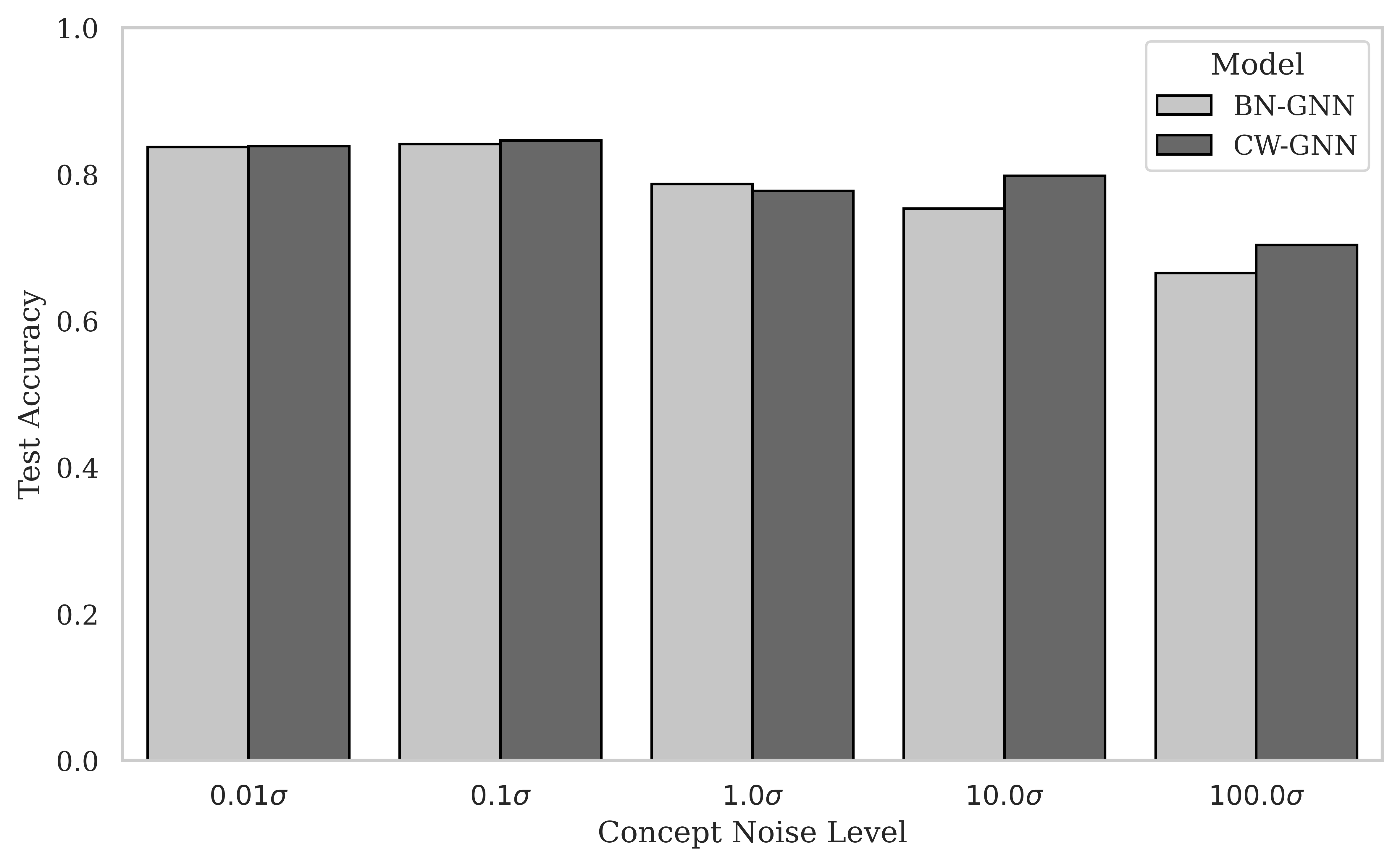}
    \caption{Bar plot of final validation accuracy for BN-GNN and CW-GNN models under varying levels of added Gaussian concept noise, expressed as multiples of the standard deviation ($n \times \sigma$).}
    \label{fig:SI_benchmarking_concept_noise_val_acc}
\end{figure}

\begin{figure}
    \centering
    \includegraphics[width=0.8\linewidth]{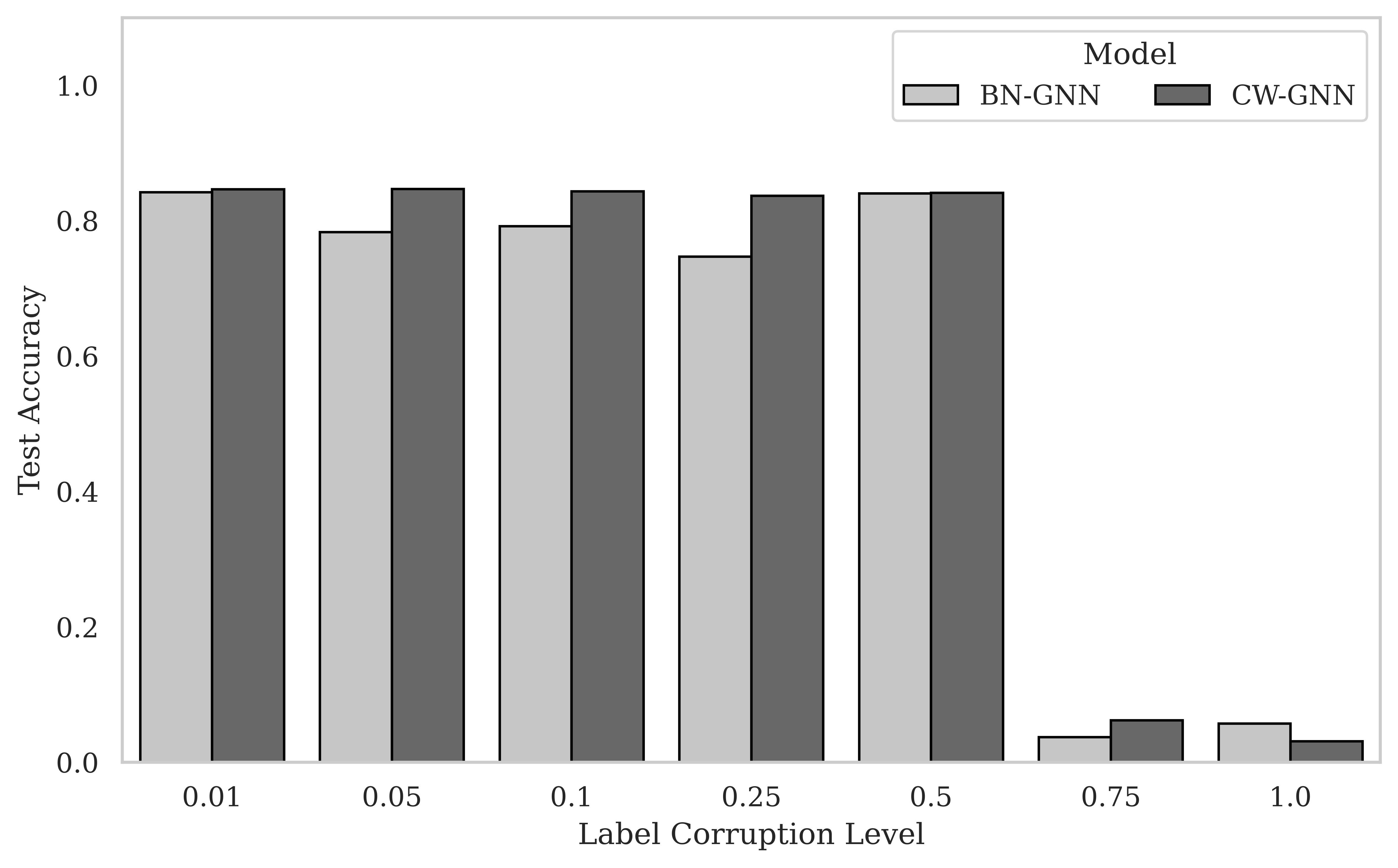}
    \caption{Bar plot comparing final validation accuracy of BN-GNN and CW-GNN models across varying levels of training set label corruption.}
    \label{fig:SI_benchmarking_label_corruption_val_acc}
\end{figure}

\begin{figure}
    \centering
    \includegraphics[width=0.8\linewidth]{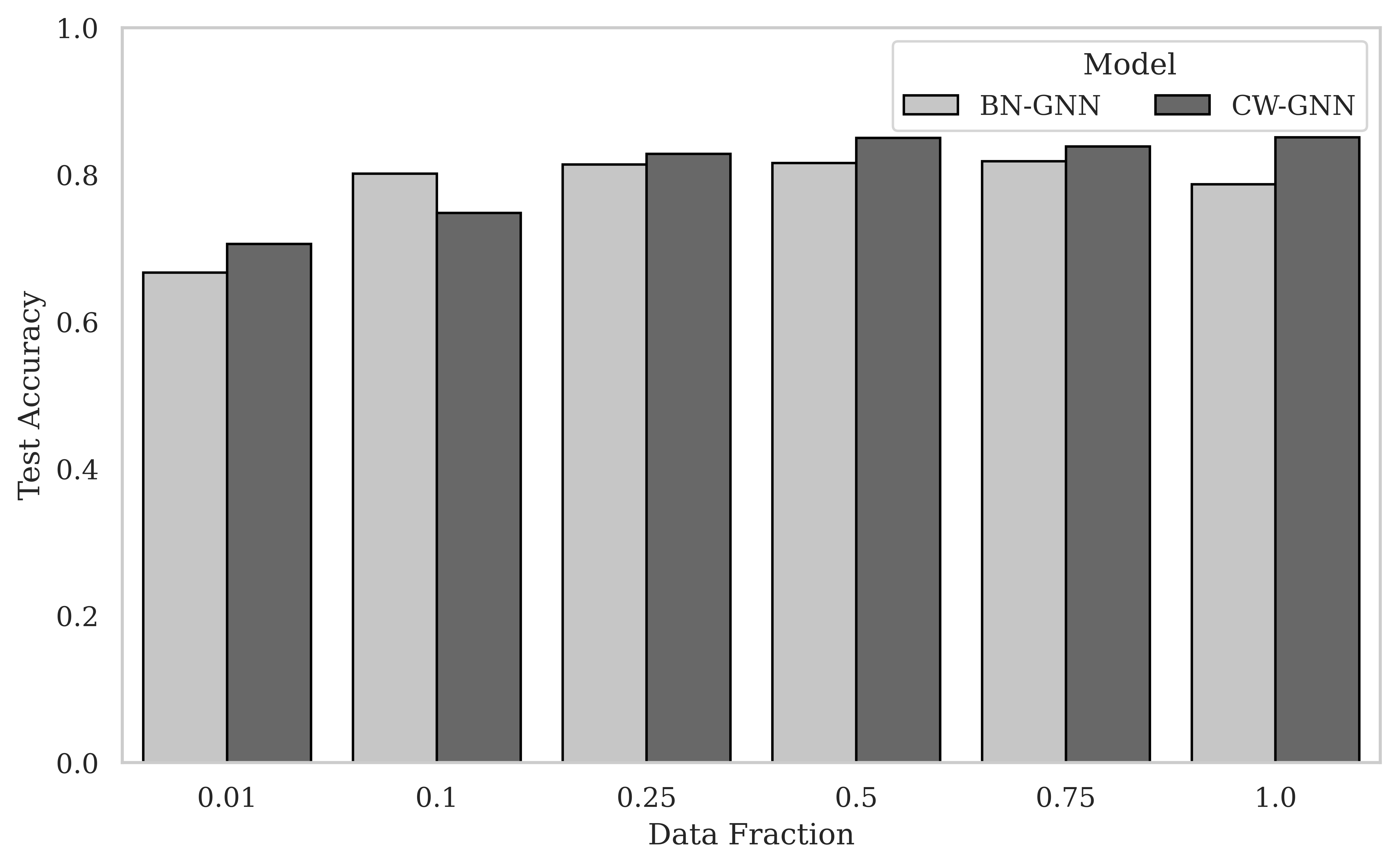}
    \caption{Bar plot comparing final validation accuracy of BN-GNN and CW-GNN models for various sizes of the overall dataset.}
    \label{fig:SI_benchmarking_data_fraction_val_acc}
\end{figure}

The benchmarking results comparing the CW-GNN versus BN-GNN discussed in Section~\ref{sec:results_and_discussion_benchmarking} of the main text are summarized in Figures~\ref{fig:SI_unoptimized_structure_mag_clf}, \ref{fig:SI_benchmarking_concept_noise_val_acc}, \ref{fig:SI_benchmarking_label_corruption_val_acc}, and \ref{fig:SI_benchmarking_data_fraction_val_acc}. The figures show the final validation accuracies obtained under four controlled perturbations: variation of the concept-loss weight (Fig.~\ref{fig:SI_benchmarking_concept_weight_val_acc}), addition of Gaussian noise to concept labels (Fig.~\ref{fig:SI_benchmarking_concept_noise_val_acc}), random corruption of training labels (Fig.~\ref{fig:SI_benchmarking_label_corruption_val_acc}), and reduction of the training set size (Fig.\ref{fig:SI_benchmarking_data_fraction_val_acc}). Together, these results quantify the predictive robustness and stability of each model across sources of variability that are relevant to our numerical experiments.

\subsection{Evolution of Concept-Latent Alignment}
\begin{figure}[htbp]
    \centering
    \includegraphics[width=\linewidth]{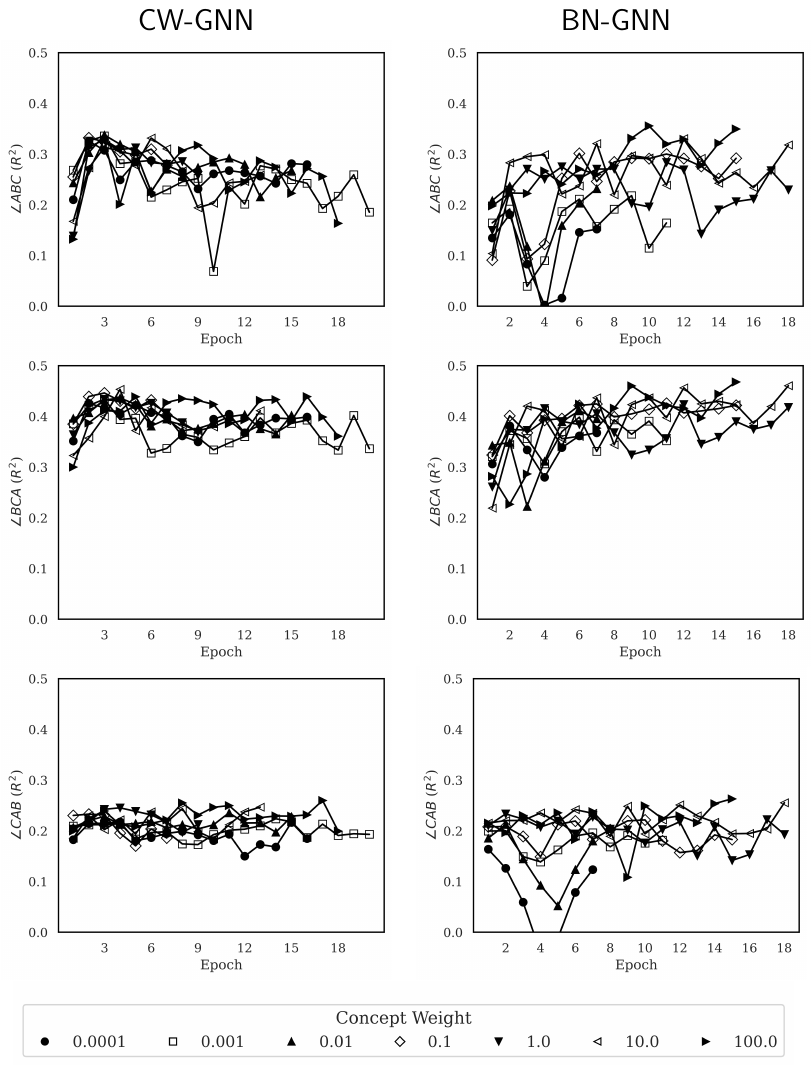}
    \caption{Learning of triplet-angle concepts in CW-GNN and BN-GNN during training under varying concept loss weights. Each row corresponds to a different angle ($A$–$B$–$C$, $B$–$C$–$A$, $C$–$A$–$B$), and columns compare the two models. Points indicate $R^2$ between model latent dimensions and the target concept at each training epoch, illustrating the models' convergence and stability in learning physically meaningful features.}
    \label{fig:SI_benchmarking_concept_weight_concepts}
\end{figure}

\begin{figure}[htbp]
    \centering
    \includegraphics[width=\linewidth]{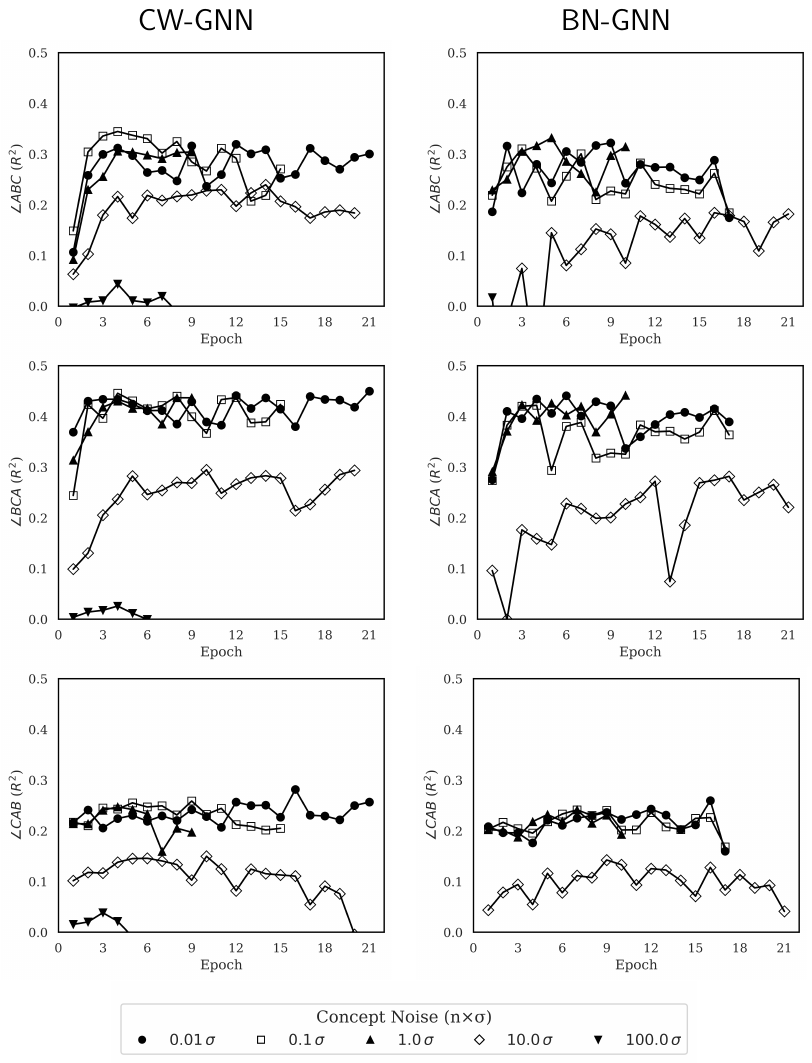}
    \caption{Learning of triplet-angle concepts in CW-GNN and BN-GNN during training under varying amounts of gaussian noise (multiples of $\sigma$) added to the training concept data. Each row corresponds to a different angle ($A$–$B$–$C$, $B$–$C$–$A$, $C$–$A$–$B$), and columns compare the two models. Points indicate $R^2$ between model latent dimensions and the target concept at each training epoch, illustrating the models' convergence and stability in learning physically meaningful features.}
    \label{fig:SI_benchmarking_concept_noise_concepts}
\end{figure}

\begin{figure}[htbp]
    \centering
    \includegraphics[width=\linewidth]{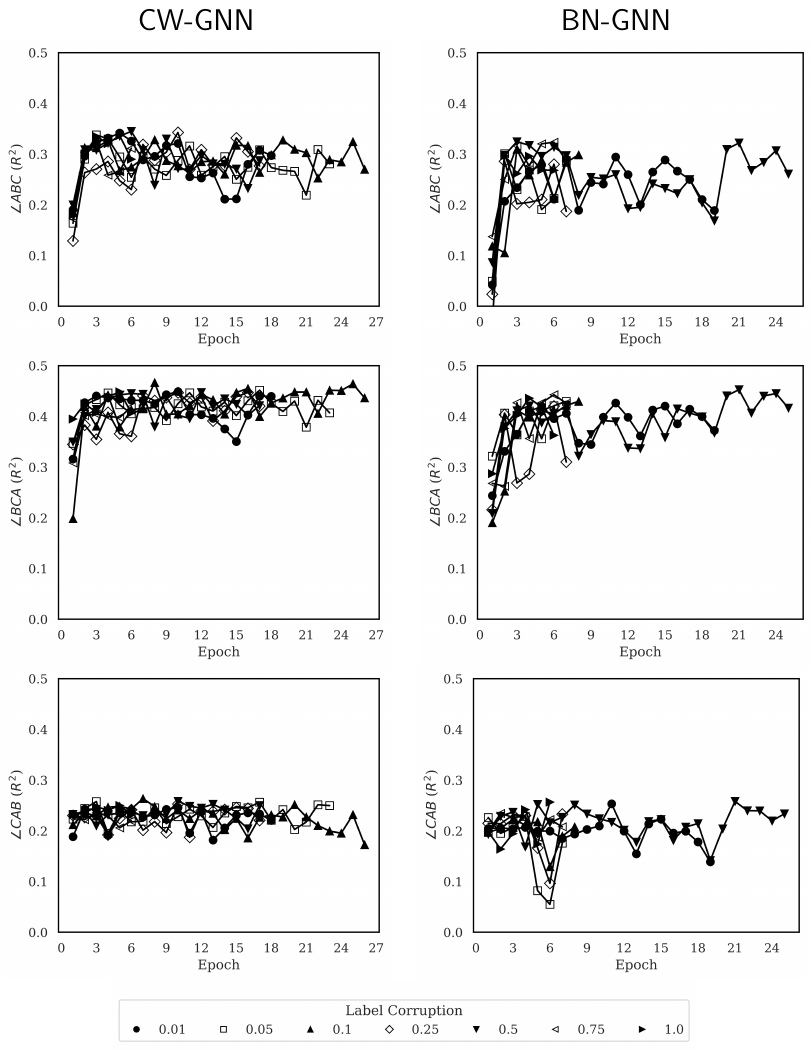}
    \caption{Learning of triplet-angle concepts in CW-GNN and BN-GNN during training under varying amounts of training data label corruption. Each row corresponds to a different angle ($A$–$B$–$C$, $B$–$C$–$A$, $C$–$A$–$B$), and columns compare the two models. Points indicate $R^2$ between model latent dimensions and the target concept at each training epoch, illustrating the models' convergence and stability in learning physically meaningful features.}
    \label{fig:SI_benchmarking_label_corruption_concepts}
\end{figure}

\begin{figure}[htbp]
    \centering
    \includegraphics[width=\linewidth]{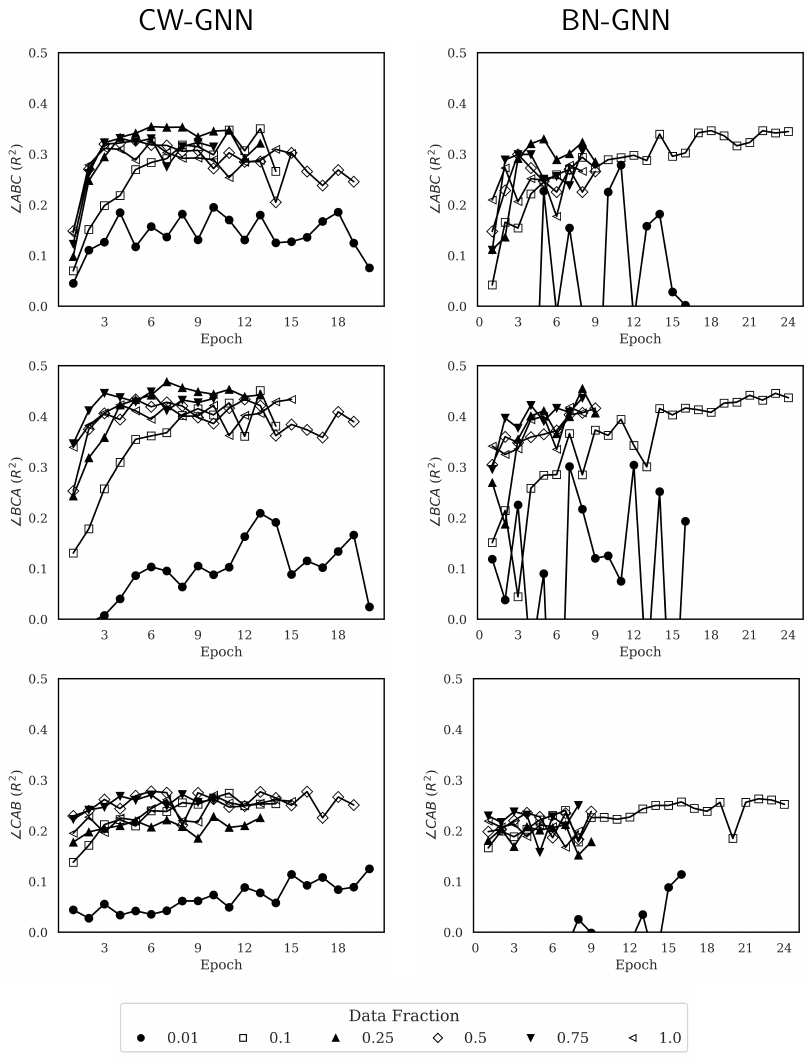}
\caption{Learning of triplet-angle concepts in CW-GNN and BN-GNN across datasets of varying size. The training, validation, and test ratios were kept constant while the total dataset size was varied. Each row shows a different angle ($A$–$B$–$C$, $B$–$C$–$A$, $C$–$A$–$B$), and columns compare the two models. Points represent the $R^2$ between model latent dimensions and the target concept at each epoch, illustrating convergence and stability in learning physically meaningful features.}
\label{fig:SI_benchmarking_data_fraction_concepts}
\end{figure}

Figures~\ref{fig:SI_benchmarking_concept_weight_concepts}, \ref{fig:SI_benchmarking_concept_noise_concepts}, \ref{fig:SI_benchmarking_label_corruption_concepts}, \ref{fig:SI_benchmarking_data_fraction_concepts} present how concept-latent alignment ($R^2$) changes during training for the same four perturbation conditions. These plots show how well each model learns and maintains alignment between latent representations and the underlying geometric concepts. Unlike BN-GNN, CW-GNN consistently forms stable, well-aligned latent subspaces, demonstrating the interpretability advantages gained through concept whitening.

Table~\ref{tab:benhmarking_model_hyperparams} summarizes the key hyperparameters and training settings for the CW-GNN and BN-GNN models used in the benchmarking experiments. Both architectures consisted of two hidden layers with mean pooling and a dropout rate of 0.2 to promote generalization. Concept Whitening (CW) or Batch Normalization (BN) was applied at the graph level depending on the model, and node features were standardized using a training-set fitted scaler. Models were trained with the Adam optimizer using a learning rate of $1\times 10^{-4}$ and a batch size of 32, and early stopping based on validation accuracy was employed to prevent overfitting. While the table provides the full details, these hyperparameter choices were guided by standard practices in graph neural network training to ensure stable and comparable performance across models. 

Table~\ref{tab:magnetic_hyperparams} summarizes the key hyperparameters used for the magnetic-ordering classification task with the CW-GNN. The model consisted of three hidden layers with intermediate dimensions chosen to balance representational capacity and computational efficiency, and mean pooling was employed to aggregate node-level features. A dropout rate of 0.2 was applied to mitigate overfitting. Training was performed with the Adam optimizer, using an initial learning rate of $1 \times 10^{-4}$ and adaptive learning rate adjustments via a ReduceLROnPlateau scheduler. Early stopping was applied to ensure convergence without unnecessary training. While the table provides the full hyperparameter details, the choices were guided by standard practices in graph neural network training and tuned to achieve stable performance on the delafossite dataset.
 
\begin{table}[htbp]
    \centering
    \caption{Hyperparameters used for magnetic property classification with CW-GNN.}
    \label{tab:magnetic_hyperparams}
    \begin{tabular}{l l}
        \hline
        \textbf{Parameter} & \textbf{Value} \\
        \hline
        Model type & CW-GNN \\
        Number of layers & 3 \\
        Hidden dimensions & [128, 256, 128] \\
        Dropout & 0.2 \\
        Pooling & Mean \\
        Learning rate & $1 \times 10^{-4}$ \\
        LR scheduler & ReduceLROnPlateau, factor=0.5, patience=5 \\
        Batch size & 32 \\
        Max epochs & 500 \\
        Early stopping patience & 20 \\
        \hline
    \end{tabular}
\end{table}

%\section*{\large Competing Interests}
%The authors declare no competing interests.

\end{document}